# A Study on the Security Requirements Analysis to build a Zero Trust-based Remote Work Environment

Haena Kim†
Korea University
haena0114@korea.ac.kr

Yejun Kim
Korea University
v3locy@korea.ac.kr

Seungjoo Kim
Korea University
skim71@korea.ac.kr

## ABSTRACT

Recently, the usage of cloud services has been increasing annually, and with remote work becoming one of the new forms of employment within enterprises, the security of cloud-based remote work environments has become important. The existing work environment relies on a perimeter security model, where accessing one's resources is based on the assumption that everything within the internal network is secure. However, due to the limitations of the perimeter security model, which assumes the safety of everything within the internal network, the adoption of Zero Trust is now being demanded. Accordingly, NIST and DoD have published guidelines related to Zero Trust architecture. However, these guidelines describe security requirements at an abstract level, focusing on logical architecture. In this paper, we conduct a threat modeling for OpenStack cloud to propose more detailed security requirements compared to NIST and DoD guidelines. Subsequently, we perform a security analysis of commercial cloud services such as Microsoft Azure, Amazon Web Service, and Google Cloud to validate these requirements. The security analysis results identify security requirements that each cloud service fails to satisfy, indicating potential exposure to threats. This paper proposes detailed security requirements based on the Zero Trust model and conducts security analyses of various cloud services accordingly. As a result of the security analysis, we proposed potential threats and countermeasures for cloud services with Zero Trust, and this is intended to help build a secure Zero Trust-based remote work environment.

## 1 INTRODUCTION

During the COVID-19 pandemic, telecommuting spread rapidly around the world, and remote work became the new normal. Although we have now entered the era of the COVID-19 endemic, remote work has become a way of life for organizations to improve employee satisfaction and boost productivity. In reference to this expansion of remote work, Microsoft CEO Satya Nadella said at the company's annual developer conference that "a digital transformation that would have taken two years has happened in two months"[1]. Gartner, a U.S. information technology research and advisory firm, forecasts that end-user spending on public cloud services worldwide will reach $597.4 billion in 2023, a 21.7% increase from 2022[2]. Traditional cloud systems are largely protected by a perimeter security model, which is why they are vulnerable to the latest reports of sophisticated cyberattacks. A Representative example is the cyberattack by the LAPTOP hacking group. The Lapsus$ hacking group, which is said to have stolen confidential information from several IT giants, including Microsoft, Nvidia, and Samsung, utilized a variety of cyberattacks. First, the group used social engineering tactics such as phishing and smishing to collect account information from employees within the companies they were targeting. The group then used the information to gain access to the company's network by disguising as normal users. The group then exploited vulnerabilities in cloud-based platforms such as Confluence,

Jira and GitLab, which are used by employees at the company for work, to gain access to confidential company information by gaining more privileges than they had at the time of access[4]. This was possible because a perimeter-based security model does not require additional authentication for previously authenticated users. In general, the perimeter-based security model divides entities accessing the system to be protected into insiders and outsiders. At this time, when an outsider attempts to access the inside of the system, authentication is performed on the outsider's identity in the perimeter-based security model. If the authentication is successful, the perimeter-based model will always trust the user. Therefore, as shown by the hacking of the Lapsus$ group, the perimeter-based security model is very limited in detecting anomalous behavior after an attacker gains access to the system. As a result, the need for a system with a Zero Trust-based security model has emerged to complement the limitations of the existing perimeter-based model. Zero Trust was first defined in 2010 in a report on Zero Trust by Forrester Research[5]. Zero Trust is the concept of "trusting no one," and the idea is that not only outsiders but also users inside a system are not trusted unconditionally, but also their identities are verified through continuous monitoring. Zero Trust has continued to be researched and developed since then, and has been standardized by organizations such as the National Institute of Standards and Technology (NIST), the Department of Defense (DoD), the National Security Agency (NSA), and the Cybersecurity and Infrastructure Security Agency (CISA). The leading guideline is "NIST Special Publication 800-207: Zero Trust Architecture," published in 2020, which provides a " Zero Trust architecture" for enterprises and federal agencies, as well as "seven principles" to guide the implementation of a Zero Trust model[6]. However, the document only describes the logical architecture at an abstract level. While the lack of specific security requirements may provide more freedom for product development, it also leaves room for developers and organizational security experts to interpret security requirements differently. This can result in inconsistent security policies being applied to systems implemented according to their own understanding. In addition, if a security incident occurs in a remote work environment, insufficient security requirements may result in a delayed and ineffective incident response. Therefore, we propose detailed security requirements for applying a Zero Trust-based security model to cloudbased remote work environments. It is important to consider security requirements when building a cloud-based remote work environment. Since the cloud environment is an environment where many users use the service at the same time, the internal resources of the enterprise must be protected from the outside. For this purpose, we selected OpenStack, which has the highest market share in Fortune 100 companies[8] among open source cloud services, and performed threat modeling. Based on the results of the threat modeling, we derived detailed security requirements for applying a Zero Trust-based security model to a cloud-based remote work environment. Based on the derived security requirements, we conducted a security analysis of Microsoft Azure, Amazon Web Service, and Google Cloud, which are known cloud services that apply the principle of Zero Trust. The security analysis in this paper was conducted by comparing the results of each cloud service analysis with the derived security requirements. As a result of the security analysis, we found that some security requirements are not satisfied in each service, which means that potential security problems may occur. Therefore, in this paper, we proposed security issues that may occur in the analyzed services and measures to mitigate them. The derived security requirements, potential threats in actual services, and mitigation measures can help developers who develop cloud services with Zero Trust to implement highly secure cloud services. This will enable cloud service providers to provide a more secure and reliable environment for users. This paper is organized as follows. Chapter 1 describes the background and necessity of this research, followed by Chapter 2 describing the Zero Trust related works, and Chapter 3 presents detailed security requirements based on Zero Trust principles by performing threat modeling for OpenStack cloud. Chapter 4 presents the security analysis results of existing commercial cloud services based on NIST and DoD security requirements and the detailed security requirements derived by this us. Finally, Chapter 5 concludes with the conclusions of this research.

## 2 RELATED WORKS

This chapter describes trends related to Zero Trust. Zero Trust-related trends explain academic trends, standardization trends, and commercial cloud trends. It also points out the limitations of current related research and describes the necessity of this study.

### 2.1 Related Academic Research Trends

In 2010, John Kindervag of Forrester Research first defined the concept of Zero Trust in "No More Chewy Centers: Introducing The Zero Trust Model Of Information Security"[5]. According to the article, a Zero Trust model is a security model that constantly verifies the identity of all entities on a network without trusting them. In 2014, Rory Ward and Betsy Beyer presented BeyondCorp: A new approach to enterprise security at the Advanced Computing Systems Association(USENIX)[9]. This was the first application of a Zero Trust-based security model to a network system proposed by Google. The security model presented in the study allows employees to work securely without the use of virtual private networks (VPNs) on untrusted networks. In 2018, Bryan Zimmer published LISA: A practical Zero Trust architecture at USENIX[11]. This is a Zero Trust-based security model adopted by Netflix, which has the advantage of providing a secure remote work environment based on simple principles. In 2020, T. Dimitrakos et al. applied the Zero Trust model to the IoT environment and studied building and maintaining trust through continuous authorization. They proposed ABAC and UCON+ based on Zero Trust, an authentication and access policy model for IoT devices in home networks, and presented a Zero Trust architecture through an access control model based on Trust Level Evaluation E-ngine (TLEE)[12]. In 2021, S. Mandal et al. proposed a cloud-based Zero Trust access control policy. In this study, a new access control policy based on Zero Trust was proposed to build a network safe from MAC spoofing attacks[13]. In 2022, P. García Teodoro et al. conducted a study on Zero Trust network access control approach based on the security profiles of users and devices. They proposed Security Attribute-based Dynamic Access Control (SADAC) as a new access control policy based on Zero Trust to enhance the security of enterprise networks and service providers, and suggested that it can improve security by minimizing risks in network and communication environments[14]. In 2023, Nisha T. N. et al. proposed a Zero Trust architecture based on three security layers related to access control and authentication by considering five security attributes and presented a six-step framework for its implementation[15]. Although there are many academic studies on Zero Trust, each study applies Zero Trust principles only to the network, access control modules in the system, etc. However, since Zero Trust architecture cannot be built as a single solution, it is not appropriate to apply Zero Trust to only some modules. Therefore, we aim to present more detailed Zero Trust-based security requirements by conducting a security analysis of existing commercial cloud services considering the overall cloud environment.

### 2.2 Standardization Trends

In 2018, the United States established the Zero Trust/SDN Steering Group to introduce the principles of Zero Trust at the federal level, and the U.S. Technology Industry Advisory Council began studying trends in Zero Trust technology[16]. As a result, in 2019, NIST recognized the importance of Zero Trust and conducted a study to develop a security model based on Zero Trust through its research institutes[17]. Subsequently, in August 2020, NIST published the Zero Trust Architecture (SP 800-207) guideline, which includes a definition of Zero Trust and the principles needed to develop it[6]. In February 2021, the Defense Information Systems Agency (DISA) and the National Security Agency (NSA) released guidance and documents related to Zero Trust. DISA and NSA published the DoD Zero Trust Reference Architecture ver 1.0 guidance, which includes the Department of Defense's information security strategy and guidance for developing and operationalizing a Zero Trust-based network security model[18]. In addition, NSA proposed the Embracing a Zero Trust Security Model which includes how to converge traditional security models with Zero Trust principles to protect assets from cyberattacks[19]. In April 2023, CISA proposed a maturity model 2.0 for zero trust with the Zero Trust Maturity Model[20]. As you can see, there is a lot of work being done to standardize Zero Trust. Since the network structure, organizational policies, and regulations of enterprises, governments, and public institutions that want to introduce Zero Trust are all different, the existing Zero Trust standards are described in the abstract to give autonomy to private organizations. However, the lack of clear and specific security requirements can make it difficult for remote work environments to be designed and operated securely, which in turn can lead to security vulnerabilities. Therefore, it is necessary to clarify and specify security requirements in order to build secure systems[21].

### 2.3 Commercial Cloud Services Trends

Microsoft has offered its cloud service, Microsoft Azure, since 2010. Microsoft Azure applies the Zero Trust principles of "verify explicitly," "use least privilege access," and "assume breach"[22]. Verifying explicitly refers to the principle that all resources and data should always be authorized and user authenticated. Least privilege access refers to the principle that all users and resources should be granted the least privileges. Breach assumption is the principle of "never trust, always verify," regardless of where a user is accessing the system and what resources they are requesting. When a user accesses the system, Microsoft Azure analyzes and collects data about possible threats to identify anomalous behavior that may occur later. Based on these principles, Microsoft applies technologies such as identity authentication, endpoint protection, application protection, data protection, infrastructure protection, network protection, and automation and orchestration to Microsoft Azure[23].

Amazon in the United States offers its cloud service, Amazon Web Service(AWS), through its data centers. AWS strictly controls access to its customers systems through its Identity and Access Management (IAM) capabilities, which include encryption and key management. AWS provides a secure network communication environment for its cloud service users through its Software Defined Perimeter(SDP) and Virtual Private Network(VPN) related features that allow cloud service users to securely use resources stored on AWS regardless of their physical location. In addition, Amazon Web Service provides cloud services based on Zero Trust, which includes five main elements: user authentication, least privilege, micro-segmentation, continuous monitoring, and cloud service automation and orchestration to ensure that cloud service users receive reliable and efficient services[24].

Google has been offering its cloud service, Google Cloud Platform, since 2011[25]. As mentioned earlier, Google has applied BeyondCorp, a Zero Trust-based security model, to its services. Through its access control features, Google Cloud Platform relies on the credentials of devices and users to access corporate resources without the use of a VPN[26]. Google provides authentication, access control, and monitoring capabilities for users and devices from a Zero Trust perspective. Related features include Identity and Access Management (IAM), Identity Aware Proxy (IAP), and Cloud IDS.

## 3 PROPOSE SECURITY REQUIREMENTS FOR ZERO TRUST-BASED CLOUD SERVICES

Threat modeling is a methodology used by many institutions and organizations to identify and manage vulnerabilities in software systems by ensuring that security is considered early in the development of software and systems. Threat modeling can identify all possible threats within a system and derive security requirements to mitigate the threats, so it has been used in several studies [27-29]. This chapter presents the security threats of cloud-based remote work environments identified by performing threat modeling. Then, we present security requirements based on the Zero Trust principle to mitigate the identified security threats. As mentioned above, we selected OpenStack as a threat modeling target and built an OpenStack-based remote work environment to perform threat modeling. We used Microsoft's STRIDE, the most popular threat modeling methodology, to analyze the security of software. This threat modeling consists of five detailed steps[30]. A brief description of the activities performed for each step is as follows.

1) Creating a data flow diagram (DFD): Create a system model by analyzing the architecture under analysis, focusing on the data flow between components.

2) Collecting Attack library: Collect all known threats to date for the components in the data flow diagram (CVE/CWE/papers/technical papers, etc.)

3) Threat analysis: Derive all potential threats that could arise from each component in the data flow diagram using the STRIDE-per-Element rule.

4) Deriving an attack tree: Derive a tree of possible attack scenarios based on the identified threats.

5) Deriving security requirements: Derive security requirements to counter attack tree sub-nodes The threat modeling results are presented in the following sections in order.

## 3.1 Data Flow Diagrams

Data flow diagrams represent the processes that are the components of a system and the data flows between them. Data flow diagrams allow you to abstract the configuration of the system under analysis from the perspective of data flow to identify the structure and attack points of the system under analysis. When creating a data flow diagram by abstracting the system to be analyzed, it is important to ensure soundness, which is the property that the analysis results should apply equally to the real system. In other words, the data flow diagram should not be missing anything compared to the analysis target. For this purpose, the data flow diagram should represent the internal functions of the module so that there is no difference between the analysis target and the abstracted model, and for this purpose, it should be written in the order of Context Level, Level 0, and Level 1. In this paper, we identified the architecture of OpenStack to understand the structure and attack points of the system under analysis[31]. Then, a data flow chart was created for the Context Level, Level 0, and Level 1 levels. The data flow diagram consists of five components: ▲ Entity, ▲Process, ▲Data Store, ▲Data Flow, and ▲Trust Boundary. The context level corresponds to the step of representing the object of analysis as a process. The context level simply shows the relationships between objects (users, OpenStack cloud services). The following Figure 1. shows the data flow diagram of the Context Level.

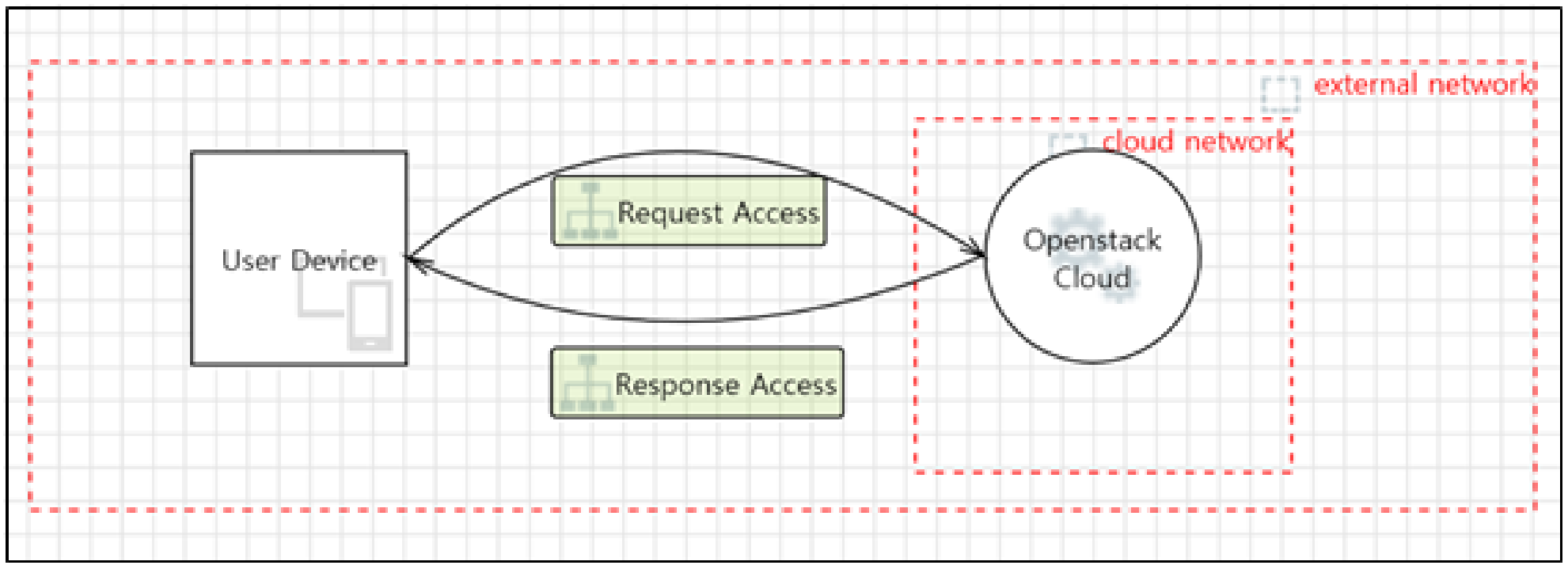

**Figure 1. Context Level DFD**

Level 0 represents the processes between major components based on OpenStack's "OpenStack Logical Architecture". Each component is further detailed in Level 1 based on OpenStack Github. The Level 0 DFD, which was created by detailing the data flow diagram of the Context Level, includes 16 components: ▲Horizon, ▲Keystone, ▲ Keystone Backend, ▲Nova, ▲Nova Data Store, ▲VM, ▲Neutron, ▲Neutron Data Store, ▲Glance, ▲Glance Data Store, ▲Swift, ▲Swift Data Store, ▲Cinder, ▲Cinder Data Store. Figure 2. shows the Level 0 data flow diagram of an OpenStack cloud.

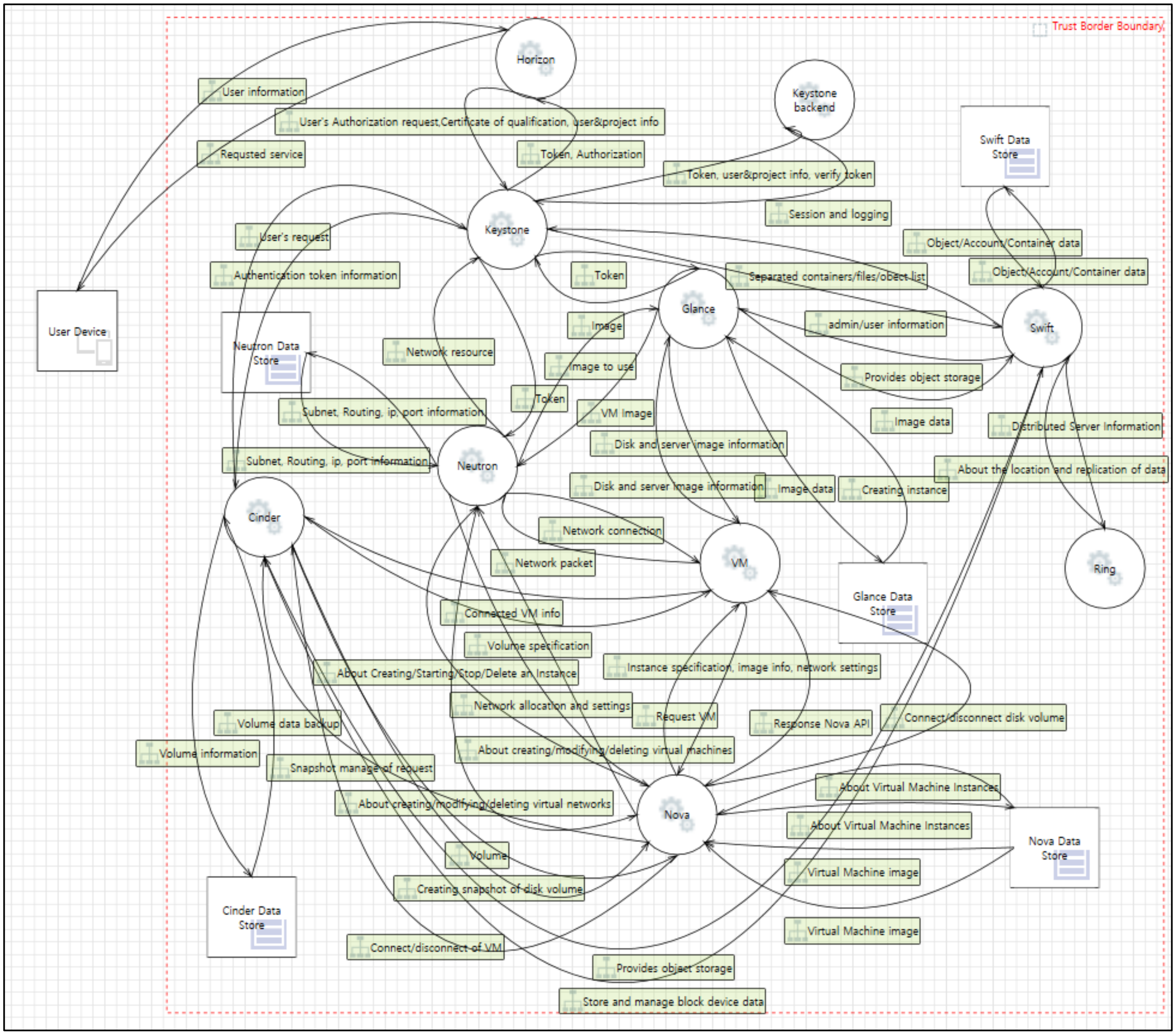

**Figure 2. Level 0 DFD**

The Level 1 data flow diagram is more detailed than the Level 0 data flow diagram in terms of how each OpenStack service component works. The Level 1 data flow diagram consists of 51 components, including Horizon, Keystone, Nova, VM, Neutron, Glance, Swift, and Cinder, which correspond to the components of the Level 0 data flow diagram. As a representative example, ▲Keystone, which is responsible for providing authentication and authorization services in OpenStack, is subdivided into ▲Keystone Backend, which supports various functions of the service ▲Identity Backend, which manages authentication for users and groups ▲Policy Backend, which manages user permissions and roles ▲Catalog Backend, which manages endpoint URLs for all OpenStack services and ▲ Token Backend, which manages temporary tokens for users. Figure 3. shows the Level 1 data flow diagram of an OpenStack cloud.

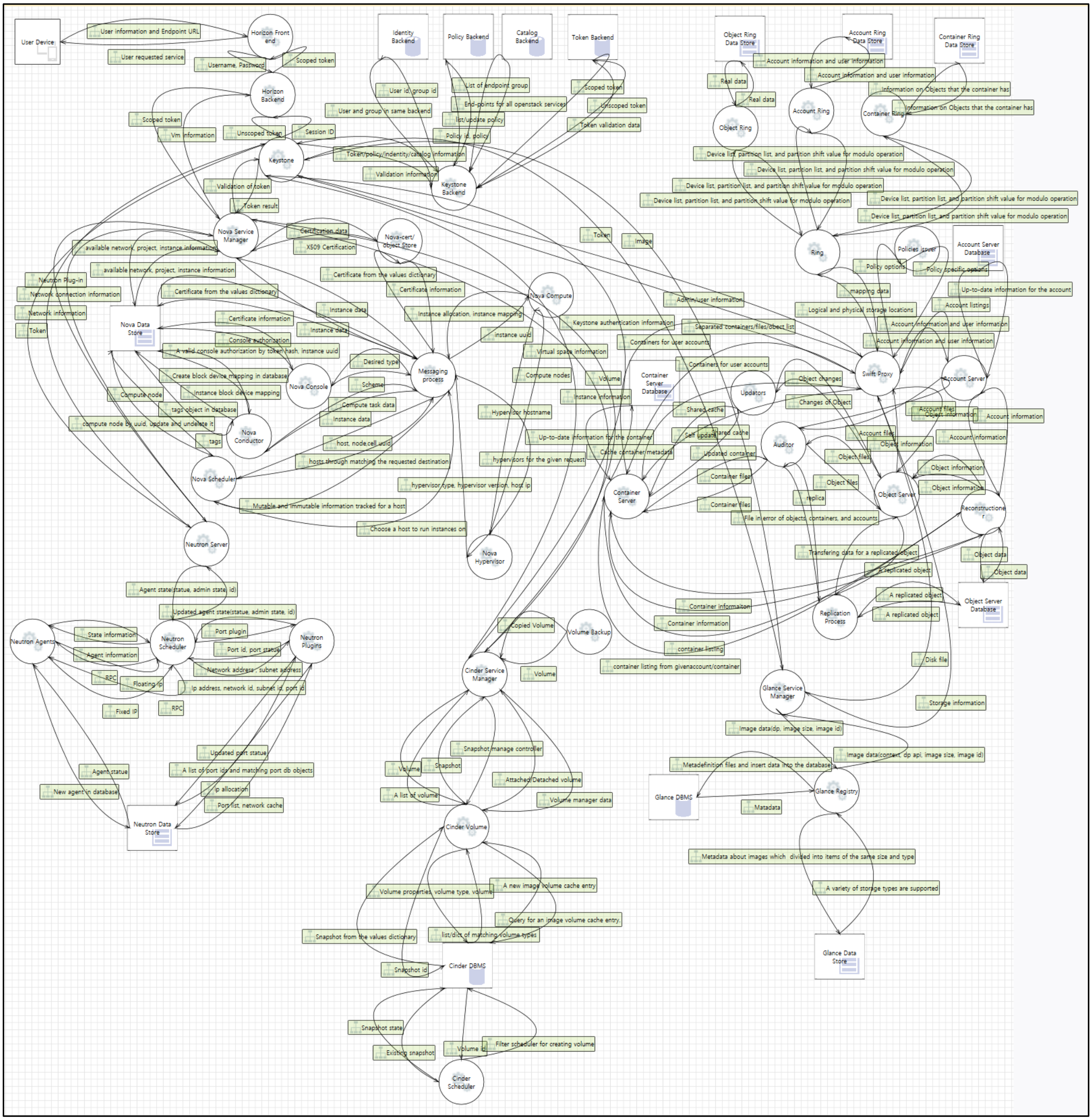

**Figure 3. Level 1 DFD**

## 3.2 Collecting Attack Libraries

The attack library collection step involves collecting all the vulnerabilities relevant to your analysis. These attack libraries collect vulnerabilities and attack techniques that are similar to the system under analysis to help you specifically identify possible threats to the system under analysis. Attack libraries are databased by collecting information on known vulnerabilities such as Common Vulnerabilities and Exposures(CVEs), Common Weakness Enumeration(CWEs), papers, standards, and technical reports. In this paper, we present 68 CVEs, 7 CWEs, and 1 paper related to OpenStack and cloud environments. A total of 569 vulnerabilities were collected from 32 publications, 4 conferences, and 458 MITRE ATT&CKs to build the attack library. The following Table 1. shows the attack libraries for the Zero Trust-based remote work environment data flow diagram derived by us, and the identifier of each attack library is represented by the abbreviation "AL" for Attack Library, an abbreviation for the source of the attack library (e.g., CVE="V", Paper="P", etc.), and a number in the order of collection, such as "AL-V-1", "AL-P-2", etc.

**Table 1. Attack Library**

| No. | Title | Ref |
|---|---|---|
| AL-C-1 | Attacking and Defending the Microsoft Cloud (Office 365 & Azure AD) | [32] |
| … | | |
| AL-V-1 | My Cloud is APT's Cloud: Investigating and Defending Office 365 | [33] |
| … | | |
| AL-A-458 | System Shutdown/Reboot | [34] |

## 3.3 Threat Analysis

The threat analysis step is to identify potential threats that may occur in the system to be analyzed based on the data flow diagram created earlier. In this paper, we used the STRIDE technique to classify threats. There are two approaches to STRIDE threat modeling: 'STRIDE per interaction' and 'STRIDE per element'. 'STRIDE per Element' has a high true positive rate that produces more results corresponding to real threats than 'STRIDE per Interaction'. Therefore, in this paper, we used the 'STRIDE per Element' approach, which performs each analysis for all components of the two approaches, and identified a total of 402 threats for each component. To verify that the identified threats are actually utilized, we include the correlation with the attack library built in Section 3.2 in Table 2.

**Table 2. STRIDE between DFD elements and Attack Library**

| Elements | Id | Name | Threat | Attack library | No. |
|---|---|---|---|---|---|
| Entity | E1 | User Device | S | AL-A-7, AL-A-20, AL-A-21, AL-A-22, AL-A-38, AL-A-39, AL-A-40, AL-A-43, AL-A-46, AL-A-59, AL-A-60, AL-A-62, AL-A-63, AL-A-64, AL-A-105, AL-W-3, AL-V-16, AL-V-17, AL-P-21, AL-A-361 | T1 |
| … | | | | | |
| Process | P43 | Volume Backup | E | AL-V-1 | T402 |

## 3.4 Attack Trees

The attack tree derivation step involves deriving attack scenarios using the threats that may occur in the system under analysis. The root node represents the attacker's attack target, and the bottom node is the entry point for executing each attack scenario. The root node represents the attacker's attack objective, and the sub node represents the entry point for executing each attack scenario. This attack tree provides a visual representation of how the threats collected in Section 3.3 can be used in an attack in practice. Based on the main threats that occur in cloud environments, we have identified three attack objectives: stealing account information stored on user devices, user denial of service, and stealing and modifying storage data. The path to achieve each attack goal is represented as a sub node, and we identified a total of 42 sub-goals through the root attack node. The first attack tree is for stealing account information stored on the user's device. An attacker can gain access to a user's device and gain privileges by sniffing the device, sending special packets, accessing through network ports, etc. After gaining user privileges, the attacker can access the organization's internal assets and steal account information on the device. The following Figure 4. shows an attack tree for stealing account information stored on a user's device.

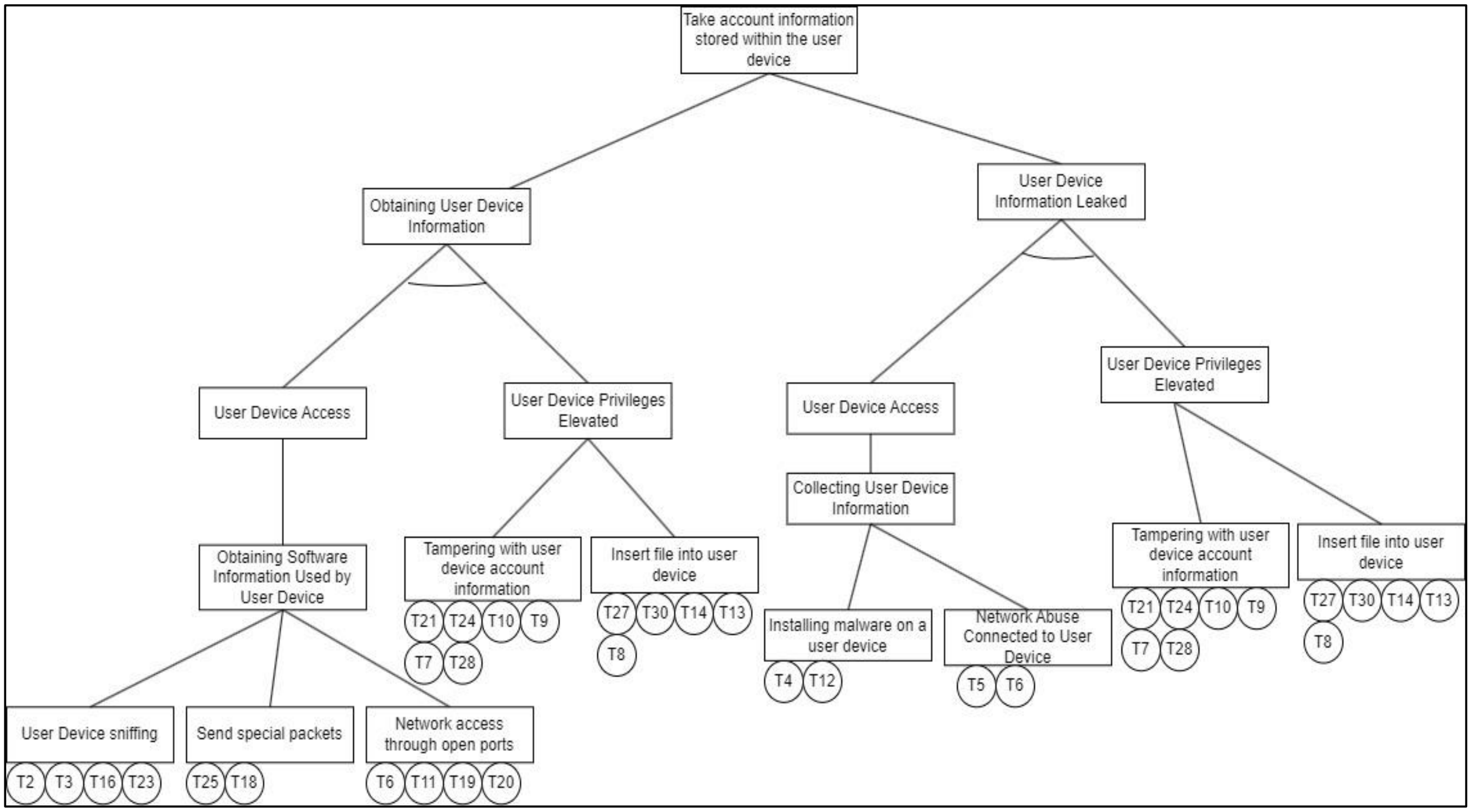

**Figure 4. Attack Tree 1**

The second attack tree is for denial of service about OpenStack services. An attacker can disrupt the use of OpenStack cloud services by destroying components in the cloud environment or generating abnormal traffic. By achieving this goal, the attacker can reduce the value of the organization by preventing users from utilizing the cloud service normally. The attack tree for denial of service consists of the denial of service root nodes of Horizon, Keystone, Nova Service Manager, Neutron server, Swift Proxy, Glance Service Manager, and Cinder Service Manager. The leaf nodes of each OpenStack service represent attack paths for an attacker to perform a denial of service attack. A malicious attacker can infect an internal employee's device and steal account information to gain access to OpenStack Cloud Horizon and gain keystone access. With Keystone access, the malicious actor can obtain an authentication token, successfully log in, and pose as a legitimate user. The attacker can then make a large number of API requests to generate dummy traffic to conduct a denial of service attack. The third attack tree is for OpenStack Storage data theft and tampering. An attacker can gain administrator privileges on OpenStack Storage and tamper with data in OpenStack Storage to disrupt user services. In addition, the attacker can perform actions to infect user devices, such as distributing malware. The following Figure 5. And Figure 6. shows an attack tree for service of denial and Data Leakage and Tampering.

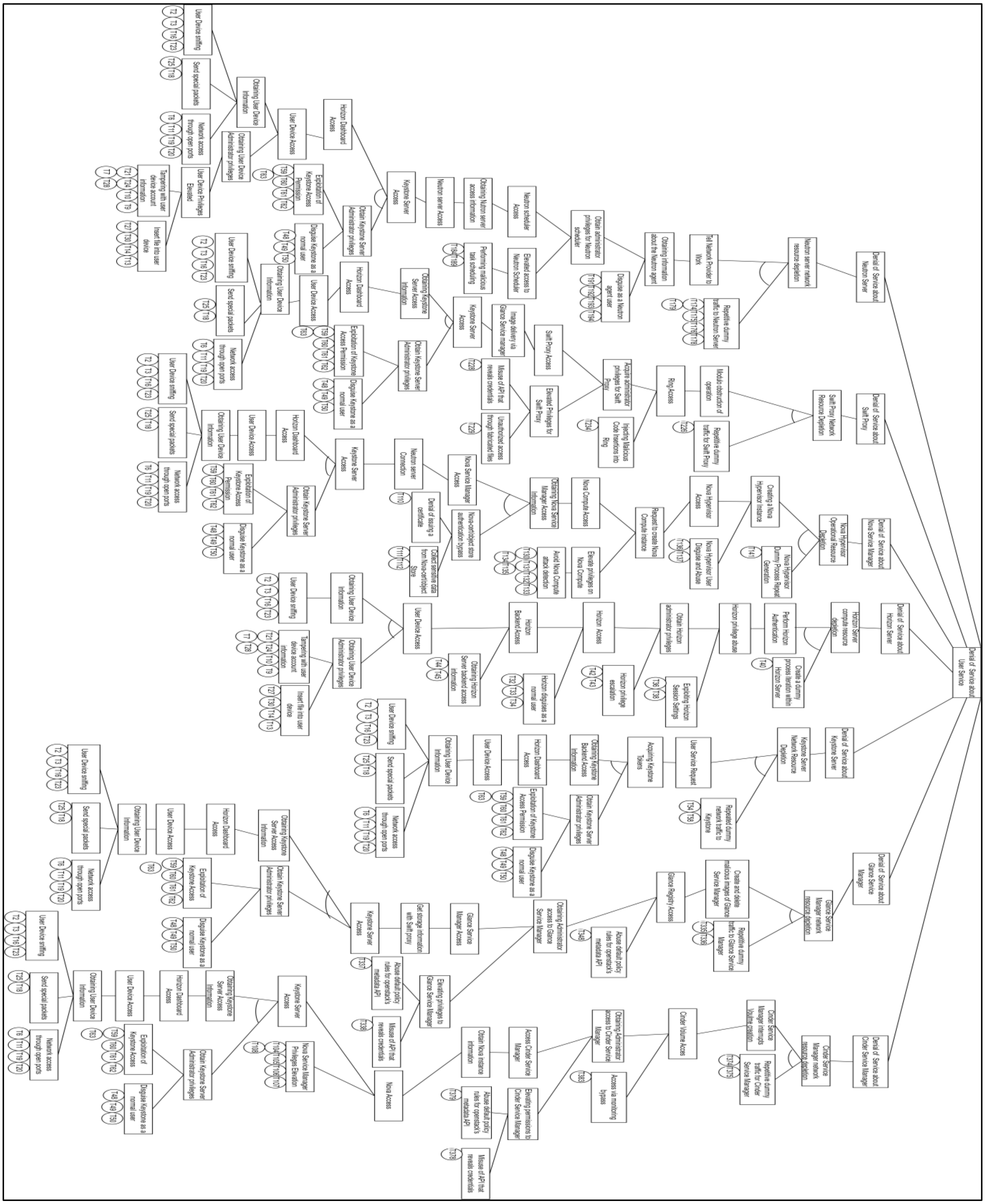

**Figure 5. Attack Tree 2**

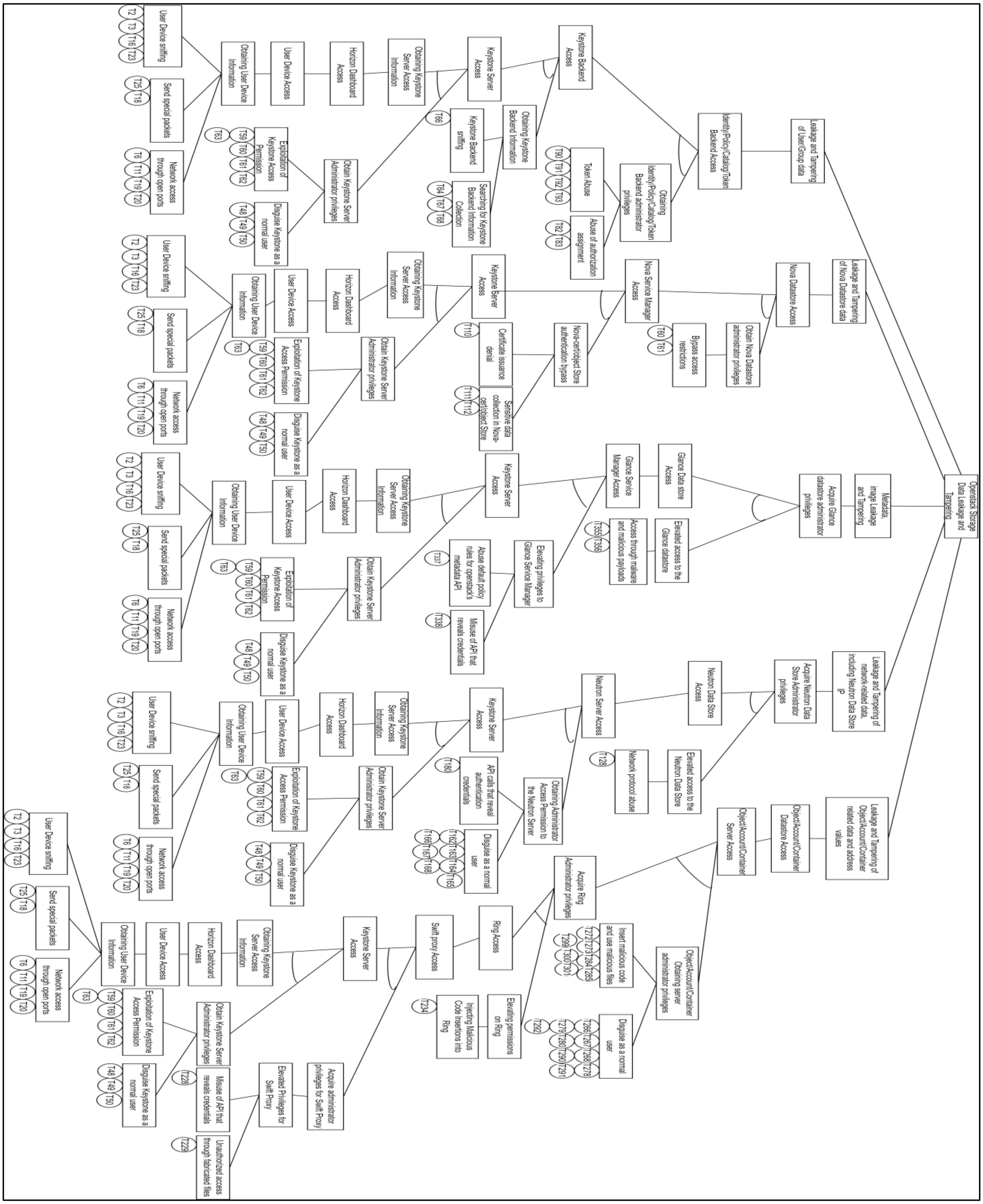

**Figure 6. Attack Tree 3**

## 3.5 Security Requirements

In this study, we derived a total of 80 security requirements as a countermeasure to the attack method at the lowest node of the attack tree so that the attacker cannot achieve the attack goal of the attack tree created in Section 3.4. OpenStack provides a total of 45 security checklists[35] for its cloud services. These checklists cover items such as setting file permissions and enabling TLS, but it is difficult to assume that Zero Trust principles have been applied, and even with these checklists, vulnerabilities are constantly being discovered[36-41]. In addition, existing academic studies that derive zero-trust-based security requirements are limited to specific situations and specific modules, not the entire cloud system. However, we derived security requirements that cover the entire cloud system. This study derives more specific security requirements than NIST and DoD because it is derived for each core component of the cloud service. The derived security requirements comply with the Zero Trust principle and are subdivided based on it, and are mainly composed of 'access control through continuous authentication', 'network encryption and segmentation', and 'continuous monitoring'. The security requirements identified by us are mapped to all seven Zero Trust principles and are refined based on NIST and DoD security requirements, NIST information system-related standards[21], and several academic studies. In general, requirements such as authentication, micro-segmentation, and traffic encryption were identified as necessary for the overall components of the cloud. For key components for cloud services, such as authentication servers, requirements such as maintaining availability were also identified as necessary. With detailed security requirements, cloud service providers can design and implement security policies for cloud-based remote work environments. This helps them prepare for threats such as unauthorized access to systems, data breaches, and malware infections. Detailed security requirements also help you identify and manage security threats. This enables you to create prevention and response strategies for potential threats. Finally, detailed security requirements help you respond quickly to security issues that arise in your system. This is because technology can be prepared in advance to effectively respond when a security issue occurs. By deriving such detailed security requirements, it is possible to strengthen the overall security of the remote work environment and minimize the damage by responding to and preventing possible security threats. Accordingly, we derived the following security requirements shown in Table 3 below to mitigate the attack tree sub nodes and prevent the attacker from executing the final attack goal.

**Table 3. Security Requirements**

| Attack/Threats | Location | Security Requirements |
|---|---|---|
| User Device Sniffing | User Device | Moving beyond existing firewalls based on ports and protocols, abnormal packets must be detected and filtered using next-generation firewalls (NGFWs) based on user ID, application recognition, and context awareness functions.<br>a) Analyze connection status, protocol header, source and destination IP, source and destination port number, TCP status, ongoing application, and if malicious packets are detected, block and record the traffic.<br>b) Utilizes iptables to block packets in transit until communication is authenticated and approved by the policy engine.<br>c) For ongoing applications, identify the application using payload size analysis, session request/response count analysis, and dynamic/regular expression pattern searching within packets.<br>d) In the case of packet encryption, Segmentation and Encapsulation is applied to segment network packets to encrypt and encapsulate each packet so that data is additionally encrypted at each step in the entire communication. |

| … |
| --- |

An example of an Attribute-Based Access Control (ABAC)-related security requirement is shown in Table 4. This example illustrates how the security requirements derived in this thesis are detailed and differentiated from those of NIST and DoD.

**Table 4. NIST, DoD and Ours Requirement Example about ABAC**

| ID | Security Requirements |
| --- | --- |
| NIST ABAC | Use Enhanced Identity Governance to requests to a policy engine service or authenticate the subject and approve the request before granting access. For this approach, enterprise resource access policies are based on **identity and assigned attributes**. The primary requirement for resource access is based on the access privileges granted to the given subject. |
| DoD ABAC | Access should be controlled by authorizing based on the **assigned attributes of the object**, the assigned attributes of the object, environmental conditions, and a set of policies specified according to those attributes and conditions. |
| Ours ABAC | Access control should be performed using attribute information (user attributes, device information, location information) through an attribute-based access control model called ABAC.<br>a) To apply ABAC, first define the attributes to control access to resources: **user roles, departments, projects, locations, and other attributes of the Keystone include entity name, domain_id, parent_id, password, project id, enabled_service, etc.).**<br>b) **Define access rules and policies** for each resource and service (ex, define policies that only users with "department" attributes "development team" and "role" is "admin").<br>C) Use Keystone's policy **engine to match attributes with policies to handle actual access requests.** Keystone's **policy engine determines whether a user or group is accessible based on the attributes.**<br>d) Modifies the "Policy configuration file" within Keystone to **enable ABAC policies and set the attributes and policies to use**. ABAC policies that allow ABAC policies to be added or changed should be tested and verified for working properly. Attempt access control with attributes and policies using Keystone API, and verify that it works as expected.<br>e) **Implement audit and monitoring of access rules assessment and execution of ABAC systems** to detect and respond to anomalies and implement automation for management and update of ABAC rules to automatically enforce rules if required. |

While NIST and DoD have published standards for zero-trust architectures, there is a lot of interpretation among cloud service providers on how to apply specific technologies. The NIST and DoD security requirements for ABAC only state that access should be controlled based on the attribute values assigned to objects. This may be insufficiently explained to cloud service providers because it does not describe what attribute values, procedures, etc. are used to perform access control. In other words, insufficiently explained security requirements do not provide in-depth understanding to cloud service providers and may cause confusion among stakeholders when building a zero-trust-based cloud environment. In addition, when security requirements are described in the abstract and contain similar content, cloud service providers cannot clearly determine which security requirements they should implement. Therefore, it is necessary for cloud service providers to implement a secure remote work environment through detailed and clear security requirements. We have detailed the security requirements into detailed implementation functions. The security requirements detailed by us specify which attributes should be applied where and how, and provide detailed procedures for applying ABAC. We believe that by complying with the detailed security requirements, cloud service providers and users will be able to keep corporate assets more secure in a remote work environment. We derived a total of 80 detailed security requirements. Table 5. shows the overall requirements we derived. The details of the security analysis performed in this research based on the detailed security requirements are described in Chapter 4.

**Table 5. A total of 80 Detailed Security Requirements**

| ID | Our Security Requirements |
|---|---|
| Ours-Secu-001 | Moving beyond traditional firewalls based on ports and protocols, next-generation firewalls (NGFWs) based on user identity, application awareness, and contextual awareness should be used to detect and filter anomalous packets.<br>a)Analyze connection status, protocol headers, source and destination IPs, source and destination port numbers, TCP status, and ongoing applications, and block and log malicious packets when they are detected.<br>b)Leverage iptables to block packets in transit until the communication is authenticated and authorized by the policy engine.<br>c) For ongoing applications, use payload size analysis, session request/response count analysis, and search for dynamic/static patterns in packets to identify applications.<br>d) For packet encryption, you can apply Segmentation and Encapsulation to further encrypt data at each step in the overall communication.<br>Segment network packets, encrypting and encapsulating each packet. |
| Ours-Secu-002 | Device Hygiene policies should be applied to keep user devices secure.<br>a) Educate your organization about security.<br>b) Identify user devices and devices and assets.<br>c)Whitelist only trusted information, rather than blocking only those that are compared to a blacklist (trusted access to the organization's network), allowing only permitted (whitelisted) users and devices to access the network.<br>d) Back up your data regularly.<br>e) Enforce security patches.<br>f) Apply continuous, automated, inventory & telemetry to locate and identify devices connected to your cloud environment, detect removals/additions, and gain an accurate picture of the full range of assets you need to monitor and protect within your enterprise. |
| Ours-Secu-003 | UEBA solutions that leverage machine learning and statistical analysis to learn behavioral patterns of users and entities should be applied.<br>a) Perform analytics on user activity in real time.<br>b) Machine learning of user activity may be performed on user devices.<br>c) Profiling and baselining: The UEBA solution builds profiles of users and entities and establishes a baseline of normal activity. This allows you to monitor behavior over a period of time and identify anomalous behavior.<br>d) Contextual analysis: Consider contextual information when analyzing user and entity behavior. For example, analyze behavior patterns in conjunction with IP address, location, time, etc. to more accurately identify anomalous |
| | behavior.<br>e) Protocol analysis: Monitor communication patterns between users and entities by analyzing network traffic and communication protocols. Identify security threats by detecting unusual protocol usage or strange data flows (manually add hosts, user IPs, host information from pre-deployed SIEM systems, scan the entire network and identify OS and other asset attributes).<br>f) Analyze user login activity: Collect and analyze user login patterns to detect anomalies.<br>g) Real-time detection and alerting: Monitor anomalies in real-time and automatically generate alerts when defined thresholds are exceeded (risk score-based profiling: Alerts are generated when the event risk score is greater than a threshold). |
| Ours-Secu-004 | Encrypted communication channels should be used for sensitive information traveling to and from your device in the cloud system.<br>a) Encrypted communication channels use encrypted means of communication, such as SSH, OpenSSH, and OpenSSL (TLS V1.3).<br>b) Cryptographic algorithms such as SEED, ARIA, and AES with key lengths of 128 bits or more, RSA with key lengths of 2048 bits or more, and SHA2 or higher are recommended. |
| Ours-Secu-005 | You should check access device policy compliance, device health, and monitor for anomalies. a) In a cloud environment, user devices can be identified as follows.<br>- Client certificates: Use client certificates to identify user devices. Grant each device a unique client certificate to access cloud services.<br>- Utilize Device Identifiers (Device IDs): Distinguish user devices by assigning a unique identifier to each device. This identifier is passed to cloud services to identify the device.<br>- Utilize IP address and location information: Utilize the IP address and location information of a user's device to identify the device. This can be used to detect access through proxies or to control access from certain geographic regions.<br>- Device biometrics: Using biometric technology (fingerprint, facial recognition, etc.) to identify user devices. Utilizing biometric features to access cloud services.<br>b) User device monitoring is performed as follows.<br>- Device profiling: Device profiling is performed by analyzing the characteristics and behavior patterns of user devices to distinguish between normal and abnormal activities. (Ex. Collect and analyze information such as device's operating system, browser version, IP address, access time, etc. to detect abnormal activities) |
| Ours-Secu-006 | Only authorized users should have access to the network, and unnecessary services and ports on network devices should be removed or blocked.<br>a) Identify and monitor assets in user devices to prevent service interruption and leakage of important information due to unauthorized access to user devices, etc.<br>- Apply continuous, automated, inventory & telemetry to locate and identify devices connected to your cloud environment, detect removals/additions, and gain an accurate picture of the full range of assets within your |
| | enterprise that need to be monitored and protected, and the ability to obtain information about these devices to monitor network access.<br>b) In a cloud environment, user devices can be identified as follows<br>- Client certificates: Use client certificates to identify user devices.<br>- Give each device a unique client certificate to access your cloud services.<br>- Leverage device identifiers (Device IDs): Distinguish user devices by assigning a unique identifier to each device.<br>- Utilize IP address and location information: Utilize the IP address and location information of a user's device to identify the device. This can be used to detect access through proxies or to control access from certain geographic regions.<br>- Device biometrics: Using biometric technology (fingerprint, facial recognition, etc.) to identify user devices. Utilizing biometric features to access cloud services.<br>b) Need to identify and categorize all assets within your cloud environment. Assets to be protected on user devices include<br>- Personal information and sensitive data: Your personal information, etc.<br>- Authentication information and credentials: Credentials stored on the device, such as login information, passwords, and authentication tokens.<br>- Work and personal files: Work documents, photos, videos, and more.<br>- Applications and software: Keep applications and software installed on your device up to date.<br>- Privacy and permission settings: Manage access to your device's camera, microphone, location information, and more.<br>- Network connections and Wi-Fi information: Manage Wi-Fi network information and network connections that your device is connected to.<br>- Security software and updates: Install security software and updates.<br>- Assess the criticality of assets: Evaluate the importance and sensitivity of each asset to determine which assets are most critical and need protection. Critical assets are secured by applying additional protective measures.<br>c) User device monitoring is performed as follows<br>- Device profiling: Device profiling is performed by analyzing the characteristics and behavior patterns of user devices to distinguish between normal and abnormal activities. (Ex. Collecting and analyzing information such as the device's operating system, browser version, IP address, and access time to detect unusual activity)<br>- Apply UEBA to analyze behavior and detect anomalies: Monitor activity on user devices in real time to detect anomalous behavior (e.g., a particular user logs in from multiple devices at the same time, or detects an unusual data transfer pattern and raises an alert). Keep devices safe by detecting malicious packets and malicious software in addition to user behavior. |
| Ours-Secu-007 | Network access and event logs of user devices should be monitored regularly.a) Log collection and aggregation settings: Allows logs generated by user devices to be sent to a centralized aggregation and storage system.<br>b) Define log data: Define what events and log data you want to collect |
| | - Select the log items that should be generated as security audit evidence when designing the system. (e.g., for user access records, access date, ID, IP address)<br>- For user access records, select the log items that should be generated as security audit evidence. Processed information subject information, tasks performed, etc. (e.g., user/user identification and authentication, administrator's management behavior, DB access, etc.)<br>c) Select an event log analysis tool: Select a tool or service to analyze log data.<br>d) Event detection and alert settings: Configure rules and notification settings to detect anomalies. For example, detect and generate alerts for failed sign.<br>-in attempts, access attempts from strange IP addresses, file access by users who don't have specific access permissions, etc.<br>e) Regular reporting and review: Regularly review and analyze log data to identify potential security events or threats. Use reports and dashboards to visually monitor and track anomalies.<br>f) Automation and response: Event log monitoring systems can incorporate automated response mechanisms to quickly respond to emergencies.<br>g) Update security policies: Update security policies and enhance security based on information gained from monitoring event logs (for example, by enforcing stronger password policies or restricting access from certain IP addresses or countries).<br>h) Learning and improvement: Analyze log data to learn from security events and develop improved security policies and processes in the future. |
| Ours-Secu-008 | The requirements of the password management procedure should be applied in the system design.<br>a) Establish rules for creating secure passwords (password complexity, length, etc.)<br>- Restrictions on the creation of passwords using easily guessed personal information such as consecutive numbers, birthdays, phone numbers, etc. and secure reissue procedures in the event of a lost or stolen user password (e.g., identity verification).<br>b) Inappropriate password types<br>- A dictionary word or combination of words<br>- Passwords that are too short or NULL (blank)<br>- A sequence of keyboard shortcuts (e.g., ABCD, QWERT, etc.)<br>- Words that can be inferred from user account information. (e.g., region, department, account name, username initials, root, rootroot, root123, admin)<br>- Passwords that utilize easy-to-guess personal information such as consecutive numbers, birthdays, phone numbers, etc.<br>c) Password complexity and length: Set a password of at least 8 characters, using a combination of letters, numbers, and special characters, that is different from your account name.<br>A combination of at least two of the following character types with a length of at least 10 characters, or a combination of at least three of the following character types with a length of at least 8 characters - Uppercase |
| | letters (26)<br>- Lowercase English letters (26)<br>- Numbers (10)<br>- Special characters (32)<br>c) Responsibility for managing passwords, including guidance on consecutive login failures, long-term inactivity (more than one year), and expiration, prohibiting automatic login, establishing secure reissuance procedures through identification in case of password loss or theft, and establishing procedures for changing characters (English case) and passwords periodically (recommended at least once a quarter, and prohibiting reuse of the same password).<br>d) Forced password change for initial access to information systems, masking during password processing (input, change). |
| Ours-Secu-009 | Data protection measures such as access control to user account information and forgery prevention should be in place. a) Require multi-factor authentication: Require multi-factor authentication (MFA) when users sign in to their account.<br>- Use two or more authentication factors (password, SMS code, fingerprint, etc.)<br>b) Role-based access control: Set up RBAC, which is strict permission-based access control for user and service accounts. Users can only get the permissions they need, which can limit access to data.<br>c) Attribute-based access control: Using attribute-based access control to determine what data a user can access. Restrict access based on attributes of a subject, resource, or environment (such as subject name, resource type, time of day, etc.).<br>d) Endpoint protection: Install endpoint security solutions on user devices to protect against malware, dangerous apps, or unauthorized access.<br>e) Data encryption: Encrypt data at rest and in transit to prevent data leaks. Data encryption and keys within cloud services. Use management solutions to enhance data protection.<br>- User account information is one of the most sensitive data in your organization or institution and is a resource that needs to be protected.<br>- A business or institution lists, categorizes, and labels account information, and encrypts it as needed.<br>- Apply encryption in transit to protect data at rest or in transit and prevent data tampering.<br>- Cryptographic algorithms use AES, DES, RSA, SHA-256, SHA-3, etc.<br>f) Detect data tampering: Detect when data is altered and perform integrity checks to determine if it has been tampered with. Generate alerts and take action on unauthorized modified data. Security information and event management (SIEM): Use SIEM tools to aggregate and analyze security logs to identify and respond to potential threats.<br>g) User education and training: Provide security education and training to users to prevent social engineering attacks and teach them how to manage and access data securely.<br>h) Updates and vulnerability management: Keep systems and software up to date and perform vulnerability management to address known security vulnerabilities. |
| | i) User activity logging: Collect and analyze logs of user activity to detect unusual activity. |
| Ours-Secu-010 | To prevent malicious file injections, ensure that only authorized users can access the internal network (service network, management network) according to the access control policy.<br>a) Micro-Segmentation: Dividing the cloud network into small subnetworks or segments, restricting traffic flow between segments and allowing access only when necessary.<br>- Block specific malicious files and ports with network packet filtering<br>b) Authentication and Multi-Factor Authentication (MFA): Users need proper authentication and authorization to access the internal network.<br>- Enforce user authentication by enforcing strong password policies and requiring multi-factor authentication (MFA) to verify a user's identity.<br>c) Endpoint protection: Install and update endpoint security solutions on user devices and servers. - Use an antivirus security program that detects and blocks malicious files and programs.<br>d) File scanning and content filtering: Perform file scanning and content filtering when uploading or downloading files to block or remediate malicious files or dangerous content.<br>e) Network Access Control (NAC): Assessing the security posture of client devices before allowing them to connect to the internal network.<br>- Keep unauthorized devices off your network.<br>f) Logging and Auditing: Log all network and user activity and review it regularly.<br>- Detect and respond to security threats based on these audit records<br>g) Policy-based access control: Set up policy-based access control and allow access only when users and devices comply with the policy.<br>h) Intelligence and threat intelligence: Update security policies and identify threats using security intelligence solutions that provide information about malicious files and threats.<br>i) User education and training: Provide security education and training to users. |
| Ours-Secu-011 | To prevent malicious file injections, you should analyze file signatures, behaviors, patterns, etc. to identify and block malicious files.<br>a) File type restrictions: Block or restrict specific file types to prevent the transfer of malicious files.<br>- Block malicious files by scanning files based on specific patterns or regular expressions in malicious files.<br>- If a malicious file does not have a known signature, analyze the file's execution behavior to detect malicious activity.<br>- Manage whitelisting to ensure only allowed software and files run and prevent malicious files from executing.b) Behavioral and pattern-based detection: Detect and block malicious files if they exhibit certain patterns or behaviors executed on the system. For example, detect anomalies such as files modifying or encrypting critical system files.<br>- Compare file hashes: Use hash values from trusted files to verify file integrity and detect changes to malicious files. |
| | c) Antivirus programs: Used in cloud environments, antivirus programs are used to identify malicious files.<br>- These programs update malicious file signatures in real time to keep up with the latest threats. |
| Ours-Secu-012 | Ensure security and system reliability through state authentication of user devices and periodic patching. For each component used in the OpenStack environment (Nova, Neutron, Cinder, etc.), apply security patches through the component's source code or deployed packages.<br>a) Service triggered or scheduled security patching: Set security patches to run automatically based on specific events or triggers in the service. Or perform patches as a regularly scheduled task.<br>- Scheduling frequency may vary depending on your environment, but once a month is recommended.<br>b) Anomaly and monitoring: Use SIEM, an anomaly and monitoring tool, to detect vulnerabilities on user devices, and identify and apply necessary patches.<br>c) Out-of-hours maintenance (Maintenance): Utilizing maintenance hours outside of key operating hours to apply patches to maintain the availability of cloud services.<br>d) Test patches in a test environment: Test and validate new security patches in a test environment in your cloud environment before applying them to your main environment.<br>e) Automatic rollback plan: Create an automatic rollback plan in case a patch breaks or has compatibility issues.<br>- Identify previous states with backups and snapshots<br>- Developing automatic rollback scripts with AutoRollback Script<br>- Set conditions to perform an automatic rollback under what circumstances. For example, perform an automatic rollback based on system performance degradation, error occurrence, etc.<br>d) Document and update: Document and update patch management processes and schedules, roles, and responsibilities to ensure efficiency and consistency in security patching. |
| Ours-Secu-013 | Dynamically configure and control network connections between users and applications to tighten access control and increase security.<br>SDP must be applied.<br>a) Adherence to the Zero Trust Principle: Requires adherence to the "do not trust" principle by default to implement SDP. Refers to the principle that all network traffic and access requests should be denied by default, and the identity of users and devices should be verified.<br>b) Verify device identification and health: To implement SDP, you need to verify device identification and health.<br>- Verify device identity, assess patch status, security software installation, malware detection, and more.<br>c) Manage user authentication and permissions: Identify network devices, device status, user IDs, and more to authorize only authorized users, and block unauthorized users entirely.<br>d) SDP implementation: SDP abstracts network connections and dynamically allows or denies connections based on the identity and state of users and devices.<br>- Evaluate network access in real time and dynamically expand or restrict access as needed.<br>e) Access policy settings: Setting access policies through SDP. These policies are dynamically adjusted based on various parameters such as user, device, application, etc. Access is denied if the required conditions are not met |
| | f) Event monitoring and logging: Continuously monitor and collect logs of events that occur through your SDP implementation. This allows you to identify and respond to access attempts and anomalies. |
| Ours-Secu-014 | Network areas should be physically or logically segregated according to cloud services, user groups, information asset criticality, legal requirements, etc.a) Requirements and assessments: Analyze your cloud environment and identify which services, groups, and information assets are critical.<br>- Check for legal requirements and compliance.<br>- Information assets can be categorized as high, medium, or low importance based on the following<br>- Business goals and values: Assess how information assets support or impact the organization's business goals and values.<br>- Confidentiality, Integrity, and Availability (CIA): Assessing the impact on the confidentiality, integrity, and availability of information assets.- Business impact analysis: Analyze the impact of a loss or breach of information assets on an organization's influence, credibility, finances, etc.<br>- Risk assessment and vulnerability analysis: Perform risk assessment and vulnerability analysis on information assets to consider acceptable risk levelsb) Plan for network segregation: Create a plan for network segregation between critical services or groups.<br>- Define how you want to implement physical or logical separation.<br>c) Virtual LANs (VLANs) or virtual networks: Implement VLANs or virtual networks for logical network separation<br>- Create isolated environments by assigning different services or groups to different VLANs.<br>d) Subnets and security groups: In a cloud environment, subnets and security groups are used to separate networks.<br>- Subnets separate networks based on IP address ranges, while security groups control traffic flow.<br>e) Set up access controls and policies: Set up access controls and policies for each isolated network zone.<br>- Access to critical information assets must be strictly controlled and policies in place to comply with legal requirements.<br>f) Security groups and firewalls: Use security groups and firewalls to control access to specific ports and protocols.<br>- Set up security groups and firewall rules for areas with sensitive information assets to protect them.<br>g) Audit and monitor: Audit and monitor network separation and security policies to identify anomalies or security threats.<br>- Improve security policies and verify compliance with audit activities.<br>h) Continuous improvement:Continually review and refine your network separation strategy as your cloud environment changes and new requirements emerge. |
| Ours-Secu-015 | An anomaly detection system for endpoints should be applied to include mobile devices, laptops, desktop PCs, servers, and hardware within the data center.<br>a) Implement an anomaly detection system: This can include SIEM systems, IDS/IPS, and machine learning/artificial intelligence-based detection systems.<br>b) Set up logs and audit trails: Configure and centralize logging of all user activity and system events to collect data for post-security incident and anomaly detection. |
| | c) Implement behavior-based analytics: Leverage machine learning and behavioral analytics to model normal user and device activity and detect anomalies.<br>d) Event response and automation: Run automated response actions when anomalies are detected and analyze security events to identify threats early.<br>User education and training: Provide security training to users and administrators to learn how to prevent social engineering attacks and comply with security policies.<br>e) Regular security assessments and audits: Periodically review your systems and security policies, and utilize the services of an external security assessment firm. |
| Ours-Secu-016 | Depending on what you're controlling access to, you'll need to specify permissions and responsibilities for access.<br>a) Identify access control areas: First, identify the various access control areas within your cloud environment.<br>- This area may vary by business process, information asset, or service.<br>b) Analyze security requirements by area: Analyze and define security requirements for each access control area.<br>- May vary based on information sensitivity, compliance requirements, and business priorities.<br>c) Define access control policies: Establish access control policies for each access control area.Policies can include user identity verification, use of multi-factor authentication (MFA), least privilege principles, etc.<br>d) Manage access rights: specify permissions and responsibilities for access based on what you want to control.- Grant only the permissions you need, and follow the principle of least privilege.<br>- Use Role-Based Access Control (RBAC) to manage permissions.<br>e) Enforce strong authentication with two-factor or MFA.<br>- Knowledge-based metrics: Answering a few secret questions only authenticated users should know.<br>- Possession-based metrics: One-time password (OTP) tokens, authentication codes, SMS codes, hardware tokens (security keys or smart cards)Inheritance-based metrics: Users use biometrics such as iris recognition, facial recognition, fingerprints, etc. Implement a multi-factor authentication process by integrating various authentication factors of your choice (for example, using a password and biometrics together, or a combination of hardware tokens and one-time verification codes to verify a user's identity).<br>d) Identify user accounts/administrator accounts and apply secure access measures.- Limited to IP and MAC addresses of accessible devices.<br>e) Regularly review access policies: Access control policies are subject to change, so keep them in line with current security requirements through regular reviews.<br>f) Access monitoring: Monitor access and collect logs to ensure compliance with access control policies. |
| Ours-Secu-017 | To control access to Keystone, you should establish formal user registration and revocation procedures and grant minimal user access based on business need.<br>a) Minimize access: The "principle of least privilege" dictates that users or systems should only be granted the privileges they need for the tasks they need to perform.<br>b) Authorization and auditing: Log and audit all access activity to track and report on who did what. |
| | c) Modernization and automation: Permissions and policies are constantly updated, and automated access control mechanisms are used to manage security policies.<br>d) Network segmentation: In a zero trust environment, segment your network to limit access to mission-critical resources.<br>e) Encryption: Encrypt data in transit and storage to prevent data exposure. |
| Ours-Secu-018 | Role-Based Access Control (RBAC), a role-based access control model, should be applied to manage user roles and privileges within an organization/enterprise. When applying RBAC, you should consider the following<br>a) Role: Define groups of users or roles. Each role represents a specific task, privilege, or set of rights associated with a job.<br>b) Permission: Defines the permissions assigned to a role. Permissions represent access to specific tasks or resources. For example, Keystone administrators can read, write, delete, etc.<br>c) User: Refers to a user who has access to the system. A user can have one or more roles, and each role can be granted different permissions.<br>d) Groups: Logically group users to make roles and permissions easier to manage.<br>e) Policy: Define a policy that represents the mapping between roles and permissions. Policies determine what permissions a user should be granted when performing certain actions.<br>f) Resource: Represents any object or data that the system wants to protect.<br>- Access to resources is determined by policy.<br>g)Access Control: Manage the relationship between roles and permissions and determine which permissions users are granted when accessing resources.<br>h) Inheritance: Define a hierarchy of roles and permissions to account for inheritance between roles (e.g., permissions from a parent role can be inherited by a child role). |
| Ours-Secu-019 | Enterprises should use just in time authorization to grant temporary access to contractors or third-party vendors when they need to perform a specific task or access a specific system.<br>a) Identity verification and authentication: Verify and authenticate the user's identity for JIT authentication.<br>- Identify users using multi-factor authentication (MFA), etc.b) Define access policies: Define access control policies for each user or role.<br>- Define the resources, services, or data that users can access.<br>c) Generate tokens on access request: When a user tries to access a particular resource, the JIT system generates an authorization token that expresses the required permissions.<br>- Apply Keystone's authentication expiration time to set a period of time to ensure that authentication occurs within a set period of time before Keystone's token is issued to access the cloud system.<br>d) Authorization token validation and authorization: The generated authorization token is submitted with the user's request. The system validates this token and grants the user the required permissions.<br>e) Manage access cycles: The JIT system monitors users' access and periodically reviews their permissions. Remove permissions when a user is no longer needed for a particular resource. |
| | - Destroying temporary accounts and replacing credentials at the end of a contract with a vendor.- When requesting access, the request includes the resource, the type of work to be performed, and the time period for which access is needed.<br>f) Anomaly detection: JIT systems monitor access activity to identify anomalies and respond immediately if necessary.<br>g) Orchestration and automation: JIT authentication systems automate the authorization and revocation process, quickly granting or removing necessary permissions.<br>h) Reporting and auditing: The JIT authentication system performs logging and auditing of access activity to verify security compliance and provide reporting. |
| Ours-Secu-020 | The cloud system should ensure accountability for all users by assigning them a uniquely identifiable identifier (token).<br>a) Assign a unique identifier: Assign a unique identifier to each user.<br>- Used to uniquely identify users within a cloud environment.<br>b) User authentication and authorization: authenticate users and grant them the appropriate permissions to access cloud resources.<br>- Permissions use Role-Based Access Control (RBAC) or Group-Based Access Control or Attribute-Based Access Control (ABAC).<br>c) Use non-guessable identifiers: Use strongly randomized identifiers(UUID or a hash value) instead of guessable identifiers (such as sequential numbers or simple usernames).<br>- Administrator and privileged accounts are restricted from using guessable identifiers (root, administrator, etc.).<br>- After installing the system, remove the manufacturer's or distributor's default account and test account, etc. or change them to an account that is difficult to guess, and establish separate control measures when using the default account.<br>d) Session management: Manage user sessions, and maintain the validity of identifiers while users access cloud resources through session management.<br>- Sessions must be strictly managed at sign-in and sign-out.<br>e) Save token issuance history: Token issuance history can be managed by saving it to a log. |
| Ours-Secu-021 | Behavioral biometrics, a method of verifying identity by monitoring a user's unique behavioral patterns, should be applied.<br>a) Data collection and analysis: Collect and analyze data on user behavior patterns.<br>- Can include keystroke patterns, typing speed, mouse movement, device acceleration, etc.<br>b) Build a machine learning model: Build a machine learning model based on the collected behavioral pattern data.<br>- Used to learn your behavior patterns and verify your identity.<br>c) Create dynamic profiles: Each user has unique behavior patterns, so create dynamic profiles for each user.<br>- Serves to describe and store user behavior patterns.<br>d) Detect and analyze behavioral patterns: Detect and analyze user behavior. patterns in real time to identify normal |
| | and anomalous behavior. |
| Ours-Secu-022 | You should apply Continuous Multifactor Authentication, which uses two or more different factors to perform authentication.<br>a) User authentication: Use MFA to authenticate users the first time they access cloud resources.Typically use two or more authentication factors. (e.g., leverage a variety of authentication factors, including email, SMS, physical security tokens, one-time passwords (OTPs), biometrics, etc.)<br>b) Session-long identity verification: Performing identity verification even after a user signs in. For example, requesting a new one-time authentication code every certain amount of time or using biometrics (fingerprint, facial recognition).<br>c) Risk-based assessment: Monitor user activity and login attempts to detect anomalies or risky activity. For example, if a login attempt occurs from a different location than before, ask the user for MFA again.<br>d) Notifications and alerts: When Continuous MFA detects an anomaly, it notifies the security team or sends an alert message to the user to request further confirmation.e) Auditing and logging: Audit Continuous MFA activity and log related events for review and reporting. |
| Ours-Secu-023 | If you access the cloud system through the internal network, you must control access only through designated terminals.<br>a) Establish access control procedures for network devices. (IP or MAC control methods, number of devices, etc.)<br>-Differentiate and manage devices through IP address ranges and subnet settings.<br>-Identification by mapping hostnames to IP addresses using the Domain Name System (DNS).<br>- Differentiate devices based on user credentials and manage permissions.<br>- Network terminals can be specific types of equipment (routers, switches, servers, etc.), and distinguish terminals based on characteristics such as network speed, bandwidth, capabilities, traffic, etc.<br>b) Modernize and regularly review usage by account.<br>- Review and update account-specific usage at regular intervals (3-6 months). Regularly review to identify and take action on inactive accounts, unnecessary permissions, etc.<br>c) Monitor usage by account with SIEM to detect and respond to unusual activity or anomalies. This allows you to quickly respond to security threats d)When implementing SIEM.<br>- Includes information such as username, IP address, MAC address, device identifier, etc.<br>- Detect anomalies by collecting information such as user login and logout activity, authentication failures, etc.<br>- Track events related to authorization by monitoring resource access events, permission changes, and more for users or entities.<br>- Monitor activity logs for users and administrators |
| Ours-Secu-024 | Network and security group settings should be utilized to implement fine-grained security policies within the virtual network.<br>a) Leverage network and security group settings.<br>b) Create networks and subnets to deploy virtual machines to |
| | - Each subnet is separated by a specific role or security level.<br>c) Create security groups and define rules to control the communication of virtual machines - Security groups define rules that can allow or deny inbound and outbound traffic.<br>d) Create a port for the virtual machine and connect to its subnet.<br>e) Create security group rules to allow or deny desired communications. Classify and process traffic based on the criteria you set.<br>- Associate the security group you created with the corresponding port or virtual machine to apply the rule. |
| Ours-Secu-025 | You need to protect workloads such as applications, servers, and containers running in cloud environments with a Cloud Workload Protection Platform (CWPP) solution.<br>a) Workload identification: Identify all workloads within the cloud environment.<br>- Determine which workloads need to be protected (e.g., application workloads Neutron server, process workloads Database query execution, user Message queue).<br>b) Select a CWPP tool: Choose a CWPP solution and integrate it into your cloud environment.<br>- Determine how you want to set security policies and rules within your CWPP tool. (e.g., control repetitive dummy traffic)<br>- Must provide workload protection, vulnerability scanning, intrusion detection, logging, and monitoring.<br>c) Workload protection: Use CWPP to protect workloads.<br>- Includes malware detection, vulnerability scanning, and enforcing default security settings for workloads.<br>d) Vulnerability Management: Use CWPP to identify and manage vulnerabilities in your workloads.<br>- Automatically apply security patches, report vulnerabilities, etc.<br>- Intrusion Detection and Prevention (IDPS): Use an intrusion detection and prevention system (IDPS) integrated into your CWPP solution to identify anomalies for your workloads and block malicious activity.<br>e) Log and event monitoring: Continuously monitor and analyze logs and events generated by your workloads.<br>- Detect and respond to security threats early.<br>f) Automation and orchestration: Set up automated responses and orchestration for security events.<br>- Quickly apply security measures and minimize human intervention. |
| Ours-Secu-026 | You must ensure accountability for all users by assigning them an identifier that uniquely distinguishes them in Neutron.<br>The identifier can be any of the following<br>a) Project or Tenant: The user belongs to one or more projects or tenants.<br>- Each project or tenant can have its own network and resources.<br>- Users can connect to a specific project or tenant to manage and use resources in that project.<br>b) User ID (User ID): Neutron assigns a User ID, which is a unique identifier for each user; this User ID can be used to distinguish and authorize specific users.<br>c) Role: Manage permissions through roles assigned to users within each project or tenant.<br>- Authorize users to perform specific actions based on their role. |
| | d) Cloud ID (Cloud ID): An additional identifier used to distinguish users in a cloud environment, typically a cloud service. Used in conjunction with the provider's authentication system. |
| Ours-Secu-027 | By utilizing the cloud environment component status (list), it is necessary to check whether equipment (intrusion detection system, etc.) for detecting abnormal traffic occurrence of the control system is established.<br>a) Set up with a network-based IDS: Neutron is the component responsible for OpenStack networking, so set up with a network based IDS.<br>- Monitor network traffic to detect intrusions on your network.<br>- Deploy IDS sensors inside or outside the cloud to monitor Neutron's network traffic and detect anomalies - Detect excessive traffic with IDS sensors.<br>- Set up to quickly generate notifications and alerts when you detect anomalies.<br>- Centrally collect and store collected logs and events related to dummy traffic from your IDS.<br>- Analyze data collected from IDS to identify and investigate anomalies and potential intrusions. |
| Ours-Secu-028 | Only authorized users should have access to the network, and unnecessary services and ports on network devices should be removed or blocked.<br>a) Set up authorization and access control: Authentication and authorization mechanisms must identify authorized users and grant them access to the network You can identify users by<br>- Network ID (network identifier): Each network created in Neutron is identified by a unique Network ID. The Network ID is used to create, modify, and delete networks.<br>- Subnet ID (Subnet Identifier): An identifier for creating and identifying subnets to manage the assignment of IP addresses within a network. Subnet IDs are used to create and manage subnets.<br>- Port ID (Port Identifier): An identifier used to manage network connections to a virtual machine or instance. The port ID is used to establish and manage the connection between the virtual machine and the network.<br>- Router ID (Router Identifier): A Neutron router is a device that enables communication between virtual networks. Each router is identified by a unique Router ID.<br>- Security Group ID (Security Group Identifier): An identifier used to set up and manage firewall rules for a virtual machine. The security group ID is used to control the security policy of the virtual machine.<br>- Floating IP: An unfixed, public IP address assigned to a network resource that is connected to an external network. Floating IPs are used to support public access of instances.<br>b) Remove unnecessary services: Remove unnecessary services and ports from network devices.c) Change settings: Change the device's settings to remove or block unnecessary services and ports.- Use the device's administrator interface or command line interface (CLI).<br>d) Log and watch settings: Configure the ability to log and watch network activity to identify and respond to anomalies.<br>- Leverage log analysis and SIEM tools to monitor activity on your network. |
| Ours-Secu-029 | To prevent Neutron from making repeated requests to create networks or subnets, delete routers, and manage databases, we need to enable fine-grained access control that is partitioned based on logical attributes. |
| | a) Network Segmentation: Logical segmentation of the network within a cloud environment or data center.<br>- Segmented networks are isolated from each other and allow communication between specific services or workloads only when necessary.<br>b) Define microsegment policies: Define policies for each segment or subnet<br>- Define what kinds of traffic are allowed or blocked from each other.<br>c) Neutron Server controls and isolates traffic to the user request queue.<br>- Create virtual networks and configure subnets using the VPC or VNet feature with Neutron Agents.<br>- Restrict traffic between different subnets and allow only necessary openings<br>d) Policies are specified based on a variety of factors, including services, ports, IP address ranges, and user roles.<br>e) Introducing a means of controlling microsegments: Introduce solutions to control and manage microsegments (e.g., Virtual Networking Platform, SDN, etc.).<br>f) Automation and orchestration: Implement automation and orchestration to avoid repetitive requests and ensure consistent microsegment policy enforcement.<br>- Quickly create, modify, and delete microsegments.<br>g) Auditing and logging: Log and audit all traffic and policy changes in the microsegment.<br>- Monitor and validate security events. |
| Ours-Secu-030 | To prevent dummy traffic, you should apply workload isolation, which segments application workloads to allow only necessary connections between processes, network traffic, and API calls.<br>a) Set up virtual networks and subnets: Use virtual networks and subnets to run workloads in an isolated environment.<br>- Place each workload on a separate network or subnet to isolate communication.<br>-Enforce isolation by allowing only necessary communications and blocking unnecessary communications.(e.g., Neutron only allows network creation and virtual machine data communications)<br>b) Resource allocation and limits: Control resource usage by setting resource allocations and limits for each workload to prevent excessive resource usage and minimize interference between workloads.<br>c) Set up logs and audit logs: Collect and analyze Neutron's logs and audit logs to monitor activity between workloads. |
| Ours-Secu-031 | Define who Nova's certificates are issued to, and record and maintain them in a form that ensures certificate issuance control, management, and traceability in the event of an accident.<br>a) Event monitoring and log analysis: Monitor and analyze logs for authentication events and certificate issuance requests in OpenStack environments to detect and act on anomalous behavior with SIEM.<br>b) Set up notifications and alarms and keep a log: Set up notification and alarm mechanisms for certificate issuance requests or changes to catalog and manage related actions when they are detected.<br>- Record access subject information (account, etc.), access time, access IP or MAC information, tasks performed (access data information, post-access activity information, etc.), etc.c) Certificate and private key protection: Securely protect and manage users certificates and private keys through a key management system (KMS). |
| | - Prevent leakage of certificates and private keys and strengthen access control. |
| Ours-Secu-032 | You must establish procedures for certificate disposal to prevent leakage of sensitive information stored in the Novacert/object Store.<br>a) Request and confirm certificate revocation: Send a certificate revocation request to your SSL/TLS certificate authority. You'll need to provide the certificate's private key and the reason for revoking the certificate.<br>b) Certificate Revocation List (CRL) renewal: The CRL must be renewed to maintain a list of revoked certificates. Clients should periodically check the<br>Download the CRL to see a list of revoked certificates.<br>c) Remove all copies of a certificate: Remove all copies of retired certificates from server and client environments.<br>d) Update the Certificate Revocation List (CRL): Maintain a list of revoked certificates by updating the CRL. Clients periodically load the CRL to check the list of revoked certificates.<br>e) Update applications and systems: Perform updates or configuration changes on cloud systems that use the certificate to ensure that the retired certificate is no longer in use.<br>f) User and administrator notifications: Notify users and administrators of certificate revocation and provide instructions on how to replace it with a new, valid certificate.<br>g) Record reasons and information for certificate revocation: Record the reasons and details for certificate revocation so that they can be verified later if needed. |
| Ours-Secu-033 | To prevent unauthorized transfers of certificates from Nova-cert/object Store, you must apply Data Loss Prevention (DLP), a feature that detects and prevents the unauthorized use and transfer of information.<br>Nova-cert stores and manages the following data<br>a) Data identification and classification: Sensitive data must be identified and categorized.<br>- SSL/TLS certificates<br>b) Set up data detection rules: Set up data detection rules to detect the movement or use of sensitive data.<br>- Includes data patterns, keywords, certificate file formats, etc.<br>c) Restrict data movement and use: Control data movement or perform monitoring via SIEM.<br>- Enforce encryption when moving data within your cloud environment, or ensure that only authorized users can access and issue certificates. |
| Ours-Secu-034 | To prevent attack detection evasion due to malware, etc., you should apply Entity and Activity Auditing to monitor the activities occurring in Nova Compute.<br>a) In order to perform Entity and Activity Auditing, you need to identify activities that may occur in Nova Compute.<br>- Manage virtual machines: Create, start, stop, pause, restart, and delete virtual machines.<br>- Resource allocation and management: Allocate and manage resources such as CPU, memory, and disk to virtual machines.<br>-Virtual network settings: Manage network connectivity for virtual machines by performing network settings, assigning IP addresses, managing virtual switches, and more.<br>- Manage virtual machine images: Upload, download, and manage virtual machine images to manage the images |
| | needed to create virtual machines.<br>- Take and manage snapshots: Take, manage, and restore snapshots of a virtual machine to save and manage the state of the virtual machine. Tune virtual machine performance: Tune the performance of a virtual machine or change resource allocation to maintain or improve optimal performance.<br>b) Audit and monitor activity: Nova Compute's activity must be audited and monitored to track all actions performed within the system. This is done by collecting and analyzing log data and audit events.<br>c) Anomaly detection: Monitor entity activity in real time to understand normal behavior patterns and detect anomalies or suspicious activity.<br>d) Threat and attack detection: Detect malicious activity or security threats and respond early to keep activities on Nova Compute secure.<br>e) Generate logs and reports: Create logs and build reports on activity in Nova Compute to record and analyze security related information.<br>f) Compliance and auditing: Audit Nova Compute's activities to ensure compliance requirements are met.<br>g) Automated alerting and response: Automatically generate alerts and notifications when anomalies or malicious activity is detected and take necessary remediation actions. |
| Ours-Secu-035 | Virtual environments should be established using software with clear provenance, distribution channels, and authorship.<br>a) Set policies and guidelines: Establish policies and guidelines that encourage cloud users to only install software from known sources.<br>- Do not install software of unknown origin, unverified freeware software, open source software, unlicensed or EOL software, etc.<br>b) Use approved images: Recommend downloading and installing virtual machine images only from approved image registries.<br>c) Validate and scan images: Before you download a virtual machine image, validate the image and perform a security scan to check for malware or threats.<br>d) Utilize updates and security tools: Perform regular updates and install security tools on virtual machines to enhance detection and blocking of malicious software.<br>e) Regular security audits and surveillance: Monitor software activity in the cloud environment in real time and detect anomalies through a SIEM system.<br>- Regularly monitor installed software in your cloud environment to detect and respond to software with unknown origins or unclear distribution channels. |
| Ours-Secu-036 | You must enforce Privileged Access Management (PAM) on cloud VMs to prevent unauthorized users from creating VMs in Nova Compute.Access must be effectively managed and monitored through PAM to ensure that only users with high-level permissions can create VMs.<br>a) Limit administrator and root access: Grant administrator or root access to cloud virtual machines only when necessary, and otherwise with minimal privileges. |
| | b) Leverage multi-factor authentication (MFA): Leverage multi-factor authentication to protect administrator access and provide an additional layer of security.- Knowledge-based metrics: Answering some secret questions that only authenticated users should know.<br>- Possession-based metrics: One-time password (OTP) tokens, authentication codes, SMS codes, hardware tokens (security keys or smart cards).<br>- Inheritance-based metrics: Users use biometrics such as iris recognition, facial recognition, fingerprints, etc.Implement a multi-factor authentication process by integrating various authentication factors of your choice (for example, using a password and biometrics together, or a combination of hardware tokens and one-time authentication codes to verify a user's identity).<br>c) System logs and auditing: Monitor access and activity on virtual machines and record logs to track who created, started, stopped, etc. instances and when.<br>d) Time-limited authorization: Allow administrator access for a limited amount of time and require re-authentication if necessary.<br>e) Access control and policy management: Leverage the cloud platform's role-based access control (RBAC) and policy management capabilities to control permissions to move virtual machines or allocate resources.<br>- Service Roles: Roles that assign some or all permissions to Nova Compute.<br>- Project and User Roles: admin, member, reader<br>- Instance and Resource Roles: Manage operations on specific virtual machines. or access to Nova resources.<br>f) Automated granting and revocation of privileges: Grant and revoke administrator access through an automated process when necessary.<br>g) Compliance and auditing: Log and audit PAM activity and meet compliance requirements.<br>h) Session monitoring and termination: Leverage the ability to monitor administrator sessions, detect unusual activity, and terminate sessions if necessary. |
| Ours-Secu-037 | Protective measures should be established and applied to the hypervisor for virtual resource management. Need to identify hypervisor users and apply API and Process Micro Segmentation to allow or block communication of API calls and processes to handle communication on remote and local systems.<br>a) Hypervisors can be identified as follows<br>- Universally Unique Identifier (UUID): Each hypervisor instance is assigned a unique UUID.<br>- MAC address (Media Access Control Address): Identification through a unique MAC address assigned to the hypervisor's network interface.<br>- Serial number: Identified by using the serial number of the server or virtual machine on which the hypervisor is installed.<br>- Host name: Identify using the name of the host machine running the hypervisor.<br>- IP address: Identified using the IP address assigned to the hypervisor.<br>- Microservice tags or labels: Use tags or labels provided by the cloud. management platform to distinguish between hypervisors. |
| | - Cloud provider's resource ID: Each cloud provider assigns a unique ID that identifies the resource, which is used to identify the hypervisor.<br>- Hypervisor version and metadata: Use the hypervisor's version and other metadata to identify it.<br>b) Apply API and Process Micro Segmentation to control the interaction between applications and processes within a network of virtual machines.<br>- Isolate and separate virtual machines: Run each virtual machine in an isolated environment and limit its interaction with other virtual machines.<br>- Separate management networks: Restrict external access by separating the hypervisor management network from the network for virtual machines. |
| Ours-Secu-038 | The system on which the hypervisor is installed should have an antivirus program installed to detect malware, and the antivirus program should be kept up to date.<br>a) System patches and updates: Perform security patches and updates to the hypervisor and host operating system on a regular basis (at least quarterly) to minimize exploitable vulnerabilities.<br>b) Hardware security: Protect the hardware of the host system running the hypervisor, and keep BIOS/UEFI settings secure to prevent unauthorized access.<br>c) Hypervisor security settings: Harden your hypervisor's security settings and disable or limit unnecessary features.<br>d) Malware detection and prevention solutions: Apply malware detection and prevention solutions to your hypervisor environment to identify and block malicious activity.<br>- Utilize a malware scanning solution to scan files and processes and identify malware.<br>- Filter malicious URLs and domains to prevent users from accessing these sites and block malicious activity.<br>- Detect behavior-based patterns, including malware signatures, to block or quarantine malware. |
| Ours-Secu-039 | For security management in the cloud environment, such as detecting the occurrence of abnormal traffic, it is necessary to establish and operate an appropriate intrusion detection system, etc. and perform monitoring of security events.<br>a) Deploy an Intrusion Detection System (IDS/IPS) or SIEM: Deploy an Intrusion Detection System (IDS) or Intrusion Prevention System (IPS) in a cloud environment.<br>- Detect network and system activity, detect malicious behavior, and respond according to set policies.<br>b) Set up rules to detect anomalous traffic and events: Set up rules in your intrusion detection system (IDS/IPS) or SIEM to detect anomalous traffic and events.<br>c) Log and event collection settings: Collect logs and events from Nova and other OpenStack services and send them to a centralized log management system.<br>d) Monitor and respond to security events: Monitor security events detected by intrusion detection systems or centralized log management systems.<br>- Receive a notification, analyze the severity of the event, and take appropriate countermeasures.<br>e) Set up automated response: Set up automated response capabilities in your intrusion detection system (IDS/IPS) |
| | or SIEM to automatically take action, such as blocking traffic or modifying firewall rules, when anomalous behavior is detected.<br>f) Regularly review and remediate: Regularly review and remediate your security event and intrusion detection system settings. Keep your systems up to date and harden your measures to counter new threats and attack techniques.<br>g) Improve learning-based detection: Build learning-based detection into your intrusion detection system (IDS/IPS) or SIEM to make it more accurate at detecting security events over time. |
| Ours-Secu-040 | Ensure that only authorized users have access to the internal network (service network, management network) according to the access control policy.<br>a) The following Nova hypervisor identifiers exist.<br>- Hostname: The computer name of the host on which the hypervisor is running.<br>- IP Address (IP Address): The IP address network identification of the hypervisor.<br>- Universally Unique Identifier (UUID): A UUID is an identifier that gives each hypervisor a unique value.<br>- MAC address (Media Access Control Address): The MAC address of the hypervisor's network interface is a physically unique value.<br>- Serial Number: A unique serial number for a piece of hardware or a virtual machine in a hypervisor.<br>- Database Identifier: An identifier used within the database that manages the hypervisor information, used to retrieve information about the hypervisor.<br>- Label or Tag: Used to assign a label or tag to a hypervisor to group or identify it.<br>- Other user-defined properties: Utilize other user-defined attributes or metadata to identify the hypervisor as needed.<br>b) Allow only authorized users to access the network, and remove or block services and traffic that are unnecessary to the Nova hypervisor.<br>c) Service profiling and minimization: Profile the services running on your hypervisor and enable only the services you need.<br>- Disable or remove unnecessary services.<br>d) Network isolation: Isolate the hypervisor from other networks or use separate VLANs or subnets to restrict traffic.<br>- Block unnecessary traffic from other networks.<br>e) Configure a next-generation firewall: Configure the firewall on the hypervisor server to block unnecessary traffic.<br>- Keep only the ports you need open, and use access control lists (ACLs) to block unauthorized traffic. |
| Ours-Secu-041 | Enables fine-grained access control that splits based on logical attributes, partitioning (segregating) the network into smaller logical segments.Micro-segmentation should be applied.<br>a) Network Segmentation: First, split the virtual network into the required segments.<br>- Each segment represents an area of the network that is independent of the others, isolating traffic between virtual |
| | machines.<br>b) Configure firewall policies: Set the required security policies and firewall rules for each segment. This allows you to control traffic between segments and block unauthorized communication.<br>c) Define a microservice group: Define a microservice group by grouping related virtual machines together.<br>- Represents virtual machines that perform similar roles or functions, and restricts communication between these groups.<br>d) Use security groups: Leverage security groups in OpenStack Nova to control traffic between each microservice group - Security groups provide a mechanism to set allow/block rules for specific protocols and ports.<br>e) Use VLANs or VXLANs: Use virtualized networking technologies such as virtual LANs (VLANs) or virtual extensible LANs (VXLANs) to distinguish virtual network segments.<br>f) Encryption and authentication: Apply encryption and authentication to secure communication between microservices.<br>- Use protocols like SSL/TLS to maintain data confidentiality and enforce access controls.<br>g) Security auditing and monitoring: Even after implementing micro-segmentation, continue to perform security auditing and monitoring to detect anomalies. |
| Ours-Secu-042 | Workload isolation should be applied because it can continuously send dummy traffic for hypervisor creation.<br>a) Virtual Machine Isolation: Isolates virtual machines running within a hypervisor to provide logical isolation between each virtual machine.<br>- Each VM has an independent operating system and execution environment and should not be affected by other VMs.<br>b) Harden hypervisor security: Harden security settings for the hypervisor server to prevent illegal access and attacks from the outside.<br>- Apply updates and patches frequently to address known vulnerabilities.<br>c) Host-based firewall: Configure a host-based firewall on the hypervisor server to restrict access from the outside, controlling whether to allow access to specific ports and services.<br>d) Control communication between virtual machines: Configure firewall rules to allow communication between virtual machines only when necessary. Unnecessary network traffic should be blocked to reduce attack vectors.<br>e) Policy-based segregation: Apply policies to virtual machine placement and scheduling on Nova Hypervisor to control which different workloads run on the same host. (e.g., set policies to place workloads of the same business type on the same host and different types of workloads on separate hosts)<br>f) Physical isolation: Ensure physical isolation between hypervisor servers to block potential logical attack vectors between virtual machines.<br>g) Monitoring and auditing: Monitor virtual machine and hypervisor activity, detect and respond to anomalous behavior or access.<br>- Regular auditing and monitoring to identify changes or anomalies within the system. |
| Ours-Secu- | You should enforce Privileged Access Management (PAM) on your cloud virtual machines to prevent |
| 043 | unauthorized people from accessing the Horizon backend and attempting to successfully log in.Access must be effectively managed and monitored via PAM to ensure that only users with high-level permissions can access the Horizon backend.<br>a) Limit administrator and root access: Grant administrator or root access to the cloud Horizon backend only when necessary, and otherwise grant minimal privileges.<br>b) Leverage multi-factor authentication (MFA): Leverage multi-factor authentication to protect administrator access and provide an additional layer of security.<br>- Knowledge-based metrics: Answering a few secret questions only authenticated users should know.<br>- Possession-based metrics: One-time password (OTP) tokens, authentication codes, SMS codes, hardware tokens (security keys or smart cards).<br>Inheritance-based metrics: Users use biometrics such as iris recognition, facial recognition, fingerprints, etc. You need to implement a multi-factor authentication process by integrating the various authentication factors you choose. (For example, using a combination of passwords and biometrics, or a combination of hardware tokens and onetime verification codes to verify a user's identity.)<br>c) System logs and auditing: Monitor access activity to the Horizon backend and record logs to track who accessed it and when.<br>d) Time-limited authorization: Allow administrator access for a limited amount of time and require re-authentication if necessary.<br>e) Access control and policy management: Leverage the cloud platform's role-based access control (RBAC) and policy management capabilities to control permissions to access the Horizon backend or assign resources.<br>f) Automated granting and revocation of privileges: Grant and revoke administrator access through an automated process when necessary.<br>g) Compliance and auditing: Log and audit PAM activity and meet compliance requirements.<br>h) Session monitoring and termination: Leverage the ability to monitor administrator sessions and, if necessary, detect abnormal activity and terminate the session. |
| Ours-Secu-044 | When a user attempts to log in to Horizon, it must be controlled by secure user authentication procedures, including user authorization, limiting the number of logins, and warning of illegal login attempts.<br>a) Limit login attempts: Enforce a login failure limit feature that temporarily locks an account or blocks access after a certain number of failed login attempts.<br>b) Detect and alert on illegal login attempts: Deploy a SIEM to monitor for illegal login attempts, and when it detects anomalous login patterns, send alert messages to administrators or take security measures.<br>c) Anomalous login patterns may include the following behaviors.<br>- Many sign-in attempts: If you experience many sign-in attempts over a short period of time.<br>- Location changes: When a user's location changes rapidly and you suspect they are signing in from a different region than their physical location.<br>- Strange timezone: If a user logs in from a different timezone than the user's usual login time, the login pattern will |
| | be flagged unless it matches the user's habits.<br>d) Using different devices: If you're signing in from a different device than the user's usual device.<br>e) Sign-in frequency changes: A user's sign-in frequency suddenly increases or decreases.<br>f) Unusual sign-in location: The user is signing in from a region or country where they don't normally sign in.<br>g) Multi-account sign-in: When a user signs in with multiple accounts at the same time, or when multiple users take turns signing in with the same account. |
| Ours-Secu-045 | When a user accesses the Horizon dashboard and enters their account information, multi-factor authentication should be enforced and the account information fields should be masked.a) You should enforce strong authentication by applying two-factor or MFA.<br>- Knowledge-based metrics: Answering a few secret questions only authenticated users should know.- Possession-based metrics: One-time password (OTP) tokens, authentication codes, SMS codes, hardware tokens (security keys or smart cards).<br>Inheritance-based metrics: Users use biometrics such as iris recognition, facial recognition, fingerprints, etc. You must implement a multi-factor authentication process that integrates the various authentication factors you choose.(for example, using a password and biometrics together, or a combination of hardware tokens and one-time authentication codes to verify a user's identity).<br>b) Masking should be done when processing (inputting, changing) ID/password.<br>- Input field masking: Masks the characters entered in the username/password input field so that the actual. b) When processing (inputting, changing) ID/password, masking should be done.<br>- Input field masking: Masks the characters entered in the username/password input field to prevent the actual username/password from being exposed on the screen. Characters entered by the user are displayed as invisible characters (e.g., asterisks, circles).<br>- Input encryption: Encrypt the entered username/password before storing it in memory. |
| Ours-Secu-046 | Administrator account information in cloud systems must be managed and maintained securely.<br>a) If you have a regular user account and a root account with access to the server system, securely manage the password for the root account separately.<br>- Administrator passwords should be renewed regularly, and a password rotation policy can be applied to ensure that they change to a new password after a certain period of time (3 months, 6 months).<br>b) Keep documents or storage media that record administrator passwords confidential.<br>- The list that stores the administrator (root account) password is encrypted and stored in the Horizon backend.<br>- Give only a minimum number of people access to the admin password list.<br>- Access should be strictly controlled, with access logging and auditing mechanisms in place to monitor access through SIEM. |
| Ours-Secu-047 | All access requests should be evaluated when a user logs in, and sessions should be designed to be managed securely.<br>The design should be such that no data is shared between sessions. If duplicate logins are not allowed, a policy |
| | should be considered in the design phase that requires the creation of a new login session to either terminate a previously created login session or terminate the newly connected session.<br>a) Establish session management policies: Establish policies within the cloud system such as maximum duration of login sessions, automatic logout times, etc.<br>b) Set session expiration time: Set an expiration time for sign-in sessions to automatically log out sessions that have been inactive for a period of time. Set the session timeout to 2-5 minutes for critical functionality or 15-30 minutes for low-risk applications, and prevent new sessions from being created without the previous session being terminated.<br>c) Introduce multi-factor authentication (MFA): Introduce multi-factor authentication (MFA) at sign-in.<br>d) Risk-adaptive access control: Monitor user behavior patterns and potentially risky situations, and terminate sessions or block access when anomalies are detected.<br>e) Security logs and auditing: Create logs of login and session activity, and monitor login attempts and session activity through auditing mechanisms.<br>f) Restrict session sharing: Enhance security by limiting or blocking session sharing across devices or browsers.<br>g) Notifications and alerts: Build notification mechanisms for session activity through SIEM to make users aware of session status and alert them to anomalies.<br>h) Session token management: Prevent token leaks by ensuring secure generation, storage, and transfer of session tokens.<br>i) Periodic session-related reviews: Review session-related security policies and settings on a regular basis (at least quarterly), and make changes and improvements.<br>j) Increase user education and awareness: Users should be educated and made aware of secure session management and login practices to increase security awareness. |
| Ours-Secu-048 | Ensure that only authorized users are allowed to log in and that Openstack Horizon does not generate indiscriminate dummy traffic.<br>a) Authentication and access control: Strengthen authentication and authorization mechanisms when accessing OpenStack Horizon.<br>- User authentication is done securely using a username and password, token, or multi-factor authentication (MFA) Strong password policy: You must set a password policy to ensure that users use strong passwords.<br>b) DDoS protection: Implement a DDoS protection solution against distributed denial of service attacks (DDoS).<br>c) Web Application Firewall (WAF): Set up a WAF on OpenStack Horizon to detect and block attacks from the web application side.<br>d) Network security groups and ACLs: OpenStack uses network security groups and access control lists (ACLs) to control network traffic.<br>- Restrict access from specific ports or IP addresses if needed.<br>e) Logs and auditing: Regularly monitor logs generated by OpenStack Horizon and related services, and identify anomalies or unusual activity. |
| | f) Regular security updates: Regularly apply security updates for OpenStack and Horizon to address known security vulnerabilities.<br>g) Manage provisioning and unused resources: To avoid unnecessary dummy traffic, manage system resources efficiently and stop unused resources. |
| Ours-Secu-049 | You should utilize Horizon's interface to adjust network and security settings, create granular security groups and rules, and perform network segmentation within your cloud environment.<br>a) Create a security group: Create security groups in Horizon to control traffic from specific instances or virtual machines.<br>- Security groups are used to set access to specific ports and protocols.<br>b) Add rules: Add rules to the security group you created to control inbound and outbound traffic.<br>- Set rules to restrict access to desired ports and protocols.<br>c) Network settings: Create and manage networks and subnets in Horizon to set up a fine-grained network configuration for microsegmentation.<br>- Create virtual networks and subnets and connect instances to them.<br>d) Set up ports and connections: Use Horizon to create instances and associate them with the desired security groups.<br>- This allows you to control network traffic to specific instances based on microsegmentation rules.<br>e) Security policy management: Create and manage security policies in Horizon to centrally manage permissions and rules for each security group.<br>-Use security policies to apply consistent security rules to multiple instances or groups of networks. |
| Ours-Secu-050 | Workload Isolation should be enforced in OpenStack Horizon to allow only necessary connections between processes, network traffic, and API calls.<br>a) Create virtual networks and subnets: Create new virtual networks and subnets using Horizon.<br>- Create virtual networks and subnets that will be used to separate different workloads.<br>b) Create an instance: Creating a virtual machine or instance with Horizon. The instance you create is associated with the virtual network you created above.<br>c) Set up security groups: Set up security groups for the instances you create and control access to required ports and protocols.<br>- Separate and control traffic for each workload.<br>d) Add rules: Add the necessary inbound and outbound rules to the security group to allow or restrict communication between desired workloads.<br>- Allow only necessary traffic between workloads and maintain isolated environments.<br>e) Network connectivity and communication: Check the necessary settings to allow the instances you created to communicate over a virtual network between themselves.<br>- Configure each workload to operate within a separate virtual network.<br>f) Policy management: Use Horizon to create and manage policies to centrally manage access rights for each |
| | workload.<br>- Set and adjust policies used to control communication between workloads<br>g) Monitoring and surveillance: Monitor the communication between the created workloads and the network and detect and act on security events if necessary.<br>- Set up a watchdog dashboard with Horizon to see traffic and communication between workloads. |
| Ours-Secu-051 | Establish a procedure for checking cloud service vulnerabilities and perform the check at least once a year. a) Establish vulnerability scanning policies and procedures, including the following<br>- What to check for vulnerabilities (e.g., servers, network equipment, etc.)<br>- Vulnerability scanning frequency<br>- Assign a vulnerability scanner and accountability<br>- Vulnerability scanning procedures and methods, etc.<br>b) Conducted vulnerability scans, including the following<br>- Analyze API authentication and authorization mechanisms: Check the authentication and authorization mechanisms of Openstack APIs, and review whether user credentials are transmitted and stored securely.<br>- Enforce HTTPS usage: Protect the confidentiality and integrity of your data by encrypting all API communications over the HTTPS protocol.<br>- Validate API requests and responses: Build validation mechanisms for API requests and responses to verify external input data and prevent attacks such as unauthorized access or injection.<br>- Authentication and credential management: Securely manage users' credentials, enable them to authenticate API calls using access tokens or via two-factor or MFA authentication (e.g., tokens that can be used are OAuth or JWT), and implement multi-factor authentication processes by integrating different authentication factors of your choice (e.g., password and biometrics together, or a combination of hardware tokens and one-time verification codes to verify a user's identity).<br>- Implement credential protection: Encrypt credentials when storing them or keep them in secure storage to prevent exposure.<br>- Filter API endpoints and requests: Build filtering mechanisms for API endpoints and requests to control the exposure of unnecessary information.<br>- Access logs and auditing: Create logs of all API accesses and actions, and build auditing mechanisms to monitor access history.<br>- Manage security updates and patches: Perform regular security updates and patches of APIs and related components to remediate known vulnerabilities.<br>- Regular security reviews and assessments: Review API authentication and credentialing-related vulnerabilities on a regular basis (at least quarterly), and make fixes and improvements based on security review findings.<br>- Increase user education and awareness: Educate API users about security and reinforce the importance of protecting their credentials. |
| Ours-Secu- | To prevent API exploitation in OpenStack Swift, the following safeguards should be in place. |
| 052 | a) Enforce HTTPS usage: Protect the confidentiality and integrity of your data by encrypting all API communication within Swift over the HTTPS protocol.<br>b) Access control and authorization: Using Keystone to strengthen user authentication and authorization. Prevent unauthorized access by setting up access control and authorization for API calls appropriately.<br>c) Review API security: Determine if credentials are exposed on the API endpoint and, if necessary, revisit the API design to ensure sensitive information is not exposed.<br>d) Use API tokens: Use tokens to handle authentication and authorization when making API calls, and use tokens to avoid exposing sensitive information.<br>e) Filter API endpoints and requests: Build filtering mechanisms for API endpoints and requests to control the exposure of unnecessary information.<br>f) Security monitoring and auditing: Monitor and audit API calls and activity with SIEM to detect and respond to anomalies.<br>g) Manage security updates and patches: Regularly perform security updates and patches for Swift and related components to remediate known vulnerabilities.<br>h) Authentication and credential management: Store large data objects (files, images, backups, etc.) of users in the Swift Account server database.<br>Manage securely and protect sensitive information from being exposed when calling APIs. |
| Ours-Secu-053 | To defend against unauthorized access through manipulated files, more granular and dynamic access control, such as ABAC, should be implemented.<br>a) Define and configure policies: Identify which Attributes are used in Swift and determine under what conditions access is allowed or blocked.<br>-User attributes: user's role (e.g., Swift service manger, user, guest), location or department in the organization, and user group affiliation.<br>- Environmental attributes: Where users are connecting from (e.g., IP address, location), the current state of the system (e.g., load status, security level).<br>- Network properties: The user's network address<br>- Other security properties: how long the access control policy is valid, and whether the device or application is authorized.<br>- Identify and collect attributes: Identify and collect attributes about users, files, resources, etc. For example, define a user's role, a file's type, its creator, etc. as properties.<br>- File or resource properties: The file owner, the ownership group of the file or resource, the type of file or resource (e.g., swift image), and the sensitivity level of the file or resource.<br>b) Configure the policy engine: Configure the policy engine that manages and evaluates ABAC policies.<br>- Select a policy engine that can be used inside or outside of OpenStack Swift and click Integrations.<br>c) Define access permissions: Define which combinations of attributes allow access (for example, access can be granted if the "File Owner" attribute matches "User A" and the "File Type" attribute matches "Shareable"). |
| | d) Access request evaluation: When a user's access request comes in, it is evaluated based on attributes using the policy engine, and access is granted or denied based on ABAC rules.<br>- Requested operations (e.g., read, write, delete)<br>- Prioritization of requested actions<br>- Time zone of the requested action.e) Managing user and file properties: Building mechanisms to manage user and file properties. Ensure properties are accurate and updated.<br>f) Set up notifications and alerts: Enable ABAC policies to monitor access denied attempts and send notifications and alerts to administrators.<br>g) Security auditing and monitoring: Monitor access control and policy enforcement and perform security audits to detect and respond to anomalies. |
| Ours-Secu-054 | Swift needs to protect your virtual environment from malware such as viruses, worms, and Trojans.<br>a) Enhance overall security: Minimize vulnerabilities in your OpenStack system by securing servers and networks, managing updates, and performing vulnerability scans.<br>b) Restrict ring file access: Strictly limit access to the ring file. Set file permissions to a minimum and ensure that only necessary users have access to the ring file.<br>c) Ring file integrity checking: Build a mechanism to check the integrity of the ring file. Use hashes or digital signatures to detect and prevent tampering with the ring file.<br>d) Backing up and recovering ring files: Regularly back up your ring file and have a plan in place to restore to the backed up file.<br>- Quickly recover when ring files are corrupted or malware is injected.<br>e) Ring file monitoring and alerting: Deploy SIEM to monitor and alert on changes to address values, Object, Container, and Account related data in ring files. is detected, build a system to notify the administrator. |
| Ours-Secu-055 | You should apply the ability to segment your application workload to allow only necessary connections between components, processes, network traffic, and API calls within Swift.<br>a) Node grouping: Grouping nodes within a Swift cluster based on their roles and capabilities. For example, storage nodes and authentication nodes are categorized into different groups.<br>b) Network segregation: Configuring logical or physical network segments for each group of nodes. Traffic between different groups of nodes should be segregated, and if necessary, use a firewall to allow only allowed traffic.<br>c) Authentication and authorization: Set up authentication and authorization policies for each group of nodes in your Swift cluster.<br>- Give each group only the permissions they need, and restrict access to other groups' resources.<br>d) Service isolation: Services within Swift are separated by function. For example, a storage service and a metadata service can each run on a separate group of nodes.<br>e) Security groups and firewalls: Use security groups and firewall rules to control communication between each group of nodes. Allow only necessary services and ports, block unnecessary services.<br>f) Access control lists (ACLs): Configuring access control lists to control access to Swift objects. Grant or restrict |
| | read or write permissions to specific objects to each node group or user.<br>g) Security auditing and logging: Implement auditing and logging of all actions and events that occur within the Swift cluster. This allows anomalous behavior to be reported to the b Identify and respond to malicious activity. |
| Ours-Secu-056 | You should check whether there is equipment (intrusion detection system, etc.) to detect abnormal traffic occurrence within the OpenStack service.<br>a) Deploy IDS/IPS: Deploy an intrusion detection system on Swift to monitor network and system traffic within the cloud environment and detect anomalies.<br>- IDS detects anomalies and generates alerts; IPS detects anomalies and automatically blocks or responds to them.<br>b) Event logging and collection: Establish a system to properly collect and store security events and logs.<br>- Build an infrastructure that allows logs from components in OpenStack to be sent to a centralized log server and analyzed.<br>c) Set policies: Set policies for detecting and blocking anomalies in your IDS/IPS system. Policies can be configured based on traffic patterns, signatures, alerts, etc.<br>d) Event monitoring: Monitor and analyze security events detected by your IDS/IPS system in real time.- Quickly detect and act on anomalies, intrusion attempts, and more.<br>e) Alert and respond: Set up alerts for security events detected by the IDS/IPS system, and quickly take countermeasures. - Set up events to notify admins in real time or automatically trigger a response system if necessary.<br>f) Regular review: Regularly review and analyze security events detected by IDS/IPS systems to identify emerging threat patterns and improve response tactics.<br>g) Leverage threat intelligence: Leverage externally collected threat intelligence data to update IDS/IPS systems and improve the reliability of security events.<br>h) Integrated security monitoring: Efficiently perform security management by establishing a security information and event management (SIEM) system that can centrally monitor and manage events detected by various security solutions and security equipment, including IDS/IPS systems. |
| Ours-Secu-057 | To enable fine-grained access control within Swift, we need to apply micro-segmentation, which splits (segregates) the network into small logical segments.<br>a) Network Configuration: Implement network separation between nodes and services within a Swift cluster.<br>- Need to create logical or physical network segments to separate groups of nodes with different roles and functions.<br>b) Security groups and firewall rules: Use security groups and firewall rules to control communication between Swift nodes, ensuring that only necessary service ports are allowed. Block unnecessary ports.<br>c) Authentication and authorization at the microservice level: Implementing authentication and authorization for users and applications at the microservice level in Swift.<br>- Restrict access to users and applications that can only perform the tasks you need them to perform.<br>d) Data access control: Control access to objects within the Swift cluster |
| | - Ensure that only users or applications with authorized access can read or write data.<br>e) Security auditing and logging: Implement auditing and logging of all actions and events that occur in Swift. This allows you to identify and respond to malicious activity or strange behavior.<br>f) Update and patch management: Regularly perform updates and security patches of Swift software and related components to keep the system up to date. |
| Ours-Secu-058 | Periodically analyze security vulnerabilities for access to metadata, and develop protective measures.<br>a) Authentication and access control settings: Configure user authentication and access control using Keystone. Define which users and roles have access to the metadata API and assign them the necessary permissions. Creating, moving, deleting, copying, etc. of virtual resources must be approved by the person in charge, and the operation must be performed by an authorized administrator.<br>b) Configure API endpoint security: To secure your API endpoints, restrict external access using firewalls, security groups, etc. or configure encrypted connections using SSL/TLS.<br>c) Logging and watching API calls: Build mechanisms to log and monitor API calls and activity.<br>- Integrate log data with a centralized log server or SIEM system to detect anomalies and take protective action.<br>d) Data validation and input filtering: Implement input filtering mechanisms to validate and verify data passed to the metadata API.<br>- Prevent the entry of incorrect or malicious data (adding, modifying, or deleting metadata).<br>e) Manage security patches and updates: Regularly apply security patches and updates to Glance and related components to remediate known vulnerabilities. |
| Ours-Secu-059 | To prevent exploitation of APIs that reveal credentials in OpenStack glance, the following safeguards should be in place.<br>a) Authentication and authorization: Manage user authentication and authorization using OpenStack Keystone services.<br>- Users who want to access the API must be authenticated through appropriate authentication methods, and API access is restricted based on the roles Keystone has granted them.<br>b) API security policies and access control: Utilize network security groups and network policies to restrict access to the OpenStack Glance API.<br>- Network security groups control API access from specific IP addresses or ranges, and network policies restrict communication between specific instances.<br>c) Use API authentication tokens: Use authentication tokens on API requests to authenticate API callers and verify permissions.- Ensure that only authorized users can access your APIs.<br>d) Encryption: Protect the confidentiality of your data by utilizing encryption for API calls and data transfers.<br>- Use TLS/SSL protocol to encrypt API traffic and prevent data hijacking in the middle.<br>e) API auditing and monitoring: Set up audit logging of API calls and activity to monitor API usage and detect anomalies.<br>f) Access control and authorization: Accurately configure access control and authorization for API calls to ensure |
| | that users can only perform the actions they need to.<br>- Apply the principle of least privilege to grant only necessary permissions and limit unnecessary ones. |
| Ours-Secu-060 | You must assign an Openstack Glance identifier to ensure that only authorized users can access the internal network.<br>a) Project and user management: Create and manage projects (tenants) and users with OpenStack Keystone.<br>- Users have access to images and organize image-related resources through projects.<br>b) Define permissions and roles: Keystone uses roles to assign permissions to users. Define the permissions and roles needed for image-related tasks and assign them to users or projects.<br>c) Configure an access control policy: Configure access control policies to control access to the Glance service.<br>- Associated with roles defined in Keystone, defining which users or projects can see which images.<br>d) Control Glance API permissions: To control access to a Glance API server, set which users or projects can accept requests from that API server.<br>- Work with Keystone to authenticate and authorize users.<br>e) Management and auditing: Once you have implemented access controls, you can utilize audit logs to monitor and review all access to your systems to identify and respond to security breaches. |
| Ours-Secu-061 | To enforce workload isolation in OpenStack Glance, you need to apply methods such as running images and virtual machine instances in isolated environments and securing limited network and resource access.<br>a) Virtual Environment Isolation: Run virtual machine instances in an isolated virtualized environment.<br>- Create a virtualized environment with a hypervisor and run each instance as a separate operating system.<br>b) Security groups and network policies: Set up security groups and network policies using OpenStack Neutron.<br>- Control traffic and port access to each instance and enforce the desired level of isolation.<br>c) Micro-segmentation: Leverage OpenStack Neutron's micro-segmentation to achieve a fine-grained level of network isolation between instances. Limit the communication of each instance to increase security and maintain workload isolation.<br>d) Image validation: Validate images uploaded to Glance to ensure they don't contain malicious code, hacking tools, etc.<br>- Verify image authenticity and prevent malicious changes with image creators and signatures.<br>e) Image scanning and security checks: Scan images uploaded to Glance to detect malware and perform security checks.<br>- If necessary, block the upload of images or raise warnings based on the scan results.<br>f) Regular auditing and surveillance: Regularly audit and monitor the use of virtual machine instances and images to detect anomalous behavior and take action.<br>g) Privilege and role management: Use OpenStack Keystone to grant users only the privileges they need and manage roles to implement proper isolation and access control between workloads. |
| Ours-Secu-062 | In Openstack Nova, monitoring should be performed through SIEM to control access to unauthorized users.<br>a) Setting up instance access control policies: Configure authentication and access control policies using Keystone. |
| | - Set the required permissions to prevent unauthenticated users from performing instance-related tasks.<br>b) Nova API logging and monitoring: Log and watch Nova API requests and responses. Allows you to monitor all API activity and identify attempts by unauthorized users.<br>c) Build a SIEM system: Build a security information and event management (SIEM) system and set it up to monitor logs and events generated by the Nova API.<br>d) Anomaly detection: Set up rules and algorithms to detect anomalies using your SIEM system. (for example, configure rules to detect unauthorized user attempts to work on instances, storage operations, etc.)<br>e) Set up notifications and alerts: Enable the SIEM system to send notifications and alerts to administrators when it detects anomalies.<br>f) Access control and prevention: Take appropriate action on detected anomalies- Block access from unauthorized users or restrict storage creation and deletion without proper permissions.<br>g) Review and update your monitoring policy: Review and update your monitoring policy on a regular basis (quarterly) to ensure you can respond to new threats.<br>h) Increase user education and security awareness: Make users aware of policies and security rules for working with instances and storage and educate them. |
| Ours-Secu-063 | In case of privilege escalation of Nova Service Manager within Openstack Cinder, it is possible to create volumes without authorization based on instance information, so it is necessary to classify users who can access Cinder and grant access scope and access permissions according to each user's role.<br>a) Project/tenant and user management: Identify projects (tenants) and users (users) for which you want to define access by role. Assign the required roles for each project and user.<br>b) Define roles: Define roles and include a clear role name and description. Define if they are allowed access to create Cinder Volumes.<br>c) Privilege (Role) mapping: Mapping of roles to the privileges used by the actual OpenStack service. Define what permissions each role has.<br>d) Service and resource access permissions: Define which OpenStack services and resources you want to allow access to. (for example, specify permissions for resources such as Nova's virtual machine management, image management, network, etc.)<br>e) API access permissions: defining which API operations you want to allow. Specify which operations to allow and which to block within the Cinder, Nova APIs.<br>f) Access scope: Define the scope of what the role allows. For example, to allow access only within a specific project. |
| Ours-Secu-064 | A monitoring system should be in place within Cinder to prevent monitoring bypass.<br>a) Log collection settings: Configure log collection settings to collect log data (Cinder volume data, snapshots, and images) generated by the Cinder service.<br>b) Configure the log collection agent: Install and configure a log collection agent on each node where the Cinder service operates. The agent collects log data and sends it to your SIEM system. |
| | c) Filtering log data: SIEM systems allow you to filter collected log data to identify and analyze critical events and security issues.<br>d) Set up alerts and alarms: If necessary, configure alert and alarm settings in your SIEM system to detect security risks or unusual activity in real time and notify relevant personnel.<br>e) Report generation and dashboard configuration: SIEM systems allow you to generate reports based on log data to visualize and monitor security status. |
| Ours-Secu-065 | You need to assign an Openstack Cinder identifier so that only authorized users can access the internal network.<br>a) Manage Projects and Users: Create and Manage Projects (Tenants) and Users with OpenStack Keystone.<br>- Users have access to images and organize image-related resources through projects.<br>b) Define roles: Define roles and include a clear role name and description. Define if they are allowed access to create Cinder Volumes.<br>c) Configuring access control policies: Configure access control policies to control access to the Cinder service.<br>- Associated with roles defined in Keystone, defining which users or projects can see which images.<br>d) Control Cinder API permissions: To control access to a Cinder API server, set which users or projects can accept requests from that API server.<br>- Work with Keystone to authenticate and authorize users.<br>e) Management and auditing: Once you have implemented access controls, you can utilize audit logs to monitor and review all access to your system. |
| Ours-Secu-066 | Through Cinder's setup and management methods, you need to select the appropriate disk volumes for that virtual machine.<br>a) Create volume per virtual machine: create separate volume for each virtual machine to allocate disk space.<br>- Isolate volumes between virtual machines and protect them from conflicting with data in other virtual machines.<br>b) Perform monitoring for block storage: Perform monitoring for Cinder to manage volume delivery activity.<br>c) Access Control and Authorization: Cinder provides mechanisms for volume access control and authorization. Proper authorization and access control controls and protects access from unauthorized virtual machines or users.<br>d) Encryption: Protect virtual machine disks by encrypting them. |
| Ours-Secu-067 | When the Token Backend communicates with Keystone, important information must be sent and received through a secure channel that supports confidentiality and integrity, as data related to User id, Group id, policy, and token are exchanged.<br>a) Protect your data by enforcing SSL and SSH communications with TLS V1.3 or later, and more. |
| Ours-Secu-068 | When organizations store and encrypt sensitive data in the cloud, they should utilize key management tools to create audit logs. All events where encryption keys are generated or exchanged through the key management tool should be recorded in the audit log.<br>a) Key rotation monitoring: Also records periodic rotation events of encryption keys in the audit log. Logs information whenever a new key is generated or an existing key is exchanged.<br>b) Key usage and encryption auditing: Audit logs also record key usage and encryption behavior when encrypting |
| | sensitive data.<br>- Track what data is encrypted and what keys are used.<br>c) Access and permissions surveillance: Changes in access and permissions to key management tools are also recorded in audit logs.<br>- Tracked who accessed the key management tool and what permissions were changed.<br>d) Log policy and security: generated audit logs are securely stored in the Keystone backend.<br>- Access to logs should be limited and policies should govern retention and deletion periods.<br>e) Log analysis and monitoring: Audit logs should be monitored and analyzed in real time by the SIEM to detect evidence of unusual activity or security risks.<br>f) Log collection and aggregation: Collect and consolidate Keystone backend data generated by Keystone and store it in a centralized repository Analyze collected log data to identify normal and ideal activity.<br>- Detect security threats using techniques such as pattern analysis, behavioral analysis, and signature detection.<br>g) Notifications and alerts: Generate alerts or send notifications in real time when you detect anomalies or dangerous situations.<br>h) Threat detection and monitoring: Monitor activity in the Keystone backend in real time for efficient security event management and detect security threats based on known risk indicators or patterns. |
| Ours-Secu-069 | When a token is captured, a procedure should be established to destroy the token to prevent it from being used maliciously until it expires.<br>a) Set token expiration time: Set the expiration time for tokens in Keystone - Keystone uses a Token Revocation List (TRL) to manage token destruction<br>- TRLs must maintain a list of destroyed tokens.<br>b) Establish a token destruction management process: Establish a management process for handling token destruction in Keystone.<br>- Implement a process to add a token to the TRL when it expires or is logged out by the user.<br>c) Handle token destruction requests: Provide an API for users or administrators to make token destruction requests.<br>- Can add destroyed tokens to the TRL and invalidate them<br>d) TRL management and periodic renewal: Manage and periodically renew TRLs to remove expired tokens and keep them up to date.<br>e) Logs and audit trails: Keystone must provide logs and audit trails related to destroyed tokens so that the history of token usage and destruction can be verified. |
| Ours-Secu-070 | The Openstack Keystone Backend's token must be stored securely and access to the token must be controlled.<br>a) Store token information: Upon successful authentication, Keystone generates token information and stores it in the Token backend.- May include user ID, project ID, permissions, token validity, etc.<br>b) Retrieve and update token information: Keystone retrieves token information from the Token backend whenever it is needed to verify a user's authentication and authorization, update the status of the token, or renew the |
| | expiration period.<br>c) Enhanced token backend security: Token information in the database is encrypted with signatures and hash functions for security, or stored in a secure way.<br>- When a token expires, Keystone deletes or invalidates its information in the Token backend. This minimizes security risks and cleans up unnecessary data.<br>d) Monitoring: SIEM should be implemented within the token backend to prevent unauthorized access to token information through access control and audit trail mechanisms, and to monitor token creation and storage. |
| Ours-Secu-071 | When creating and changing user accounts in Keystone, you should establish a categorization of access rights by job function and role.<br>a) Identify jobs and roles: First, identify what jobs or roles exist within your organization. (Ex. Define the role of each of the following: system administrator, network administrator, developer, etc.)<br>b) Define and categorize privileges: Define and categorize the privileges required for each job or role. Identify which services, resources, and tasks require access and break them down into permissions.<br>c) Least privileges: Grant the least amount of privileges per role. Restrict them to perform only the necessary tasks.<br>d) Create and assign Keystone roles: Create roles defined by Keystone and assign the necessary permissions to those roles.<br>- Roles are used to control access to services and resources.<br>e) Assign roles to users: Assign each user a corresponding role. Enable users to perform tasks by granting them permissions based on the roles they need.<br>f) Set up access controls and policies: Use Keystone's Policy feature to manage roles and permissions and set up access controls.<br>- Edit policy files to control which roles can access which services and resources.<br>g) Review and modify regularly: Modify permission assignments as roles and tasks change in your organization. Regularly review users' roles and permissions and make necessary changes to maintain optimized access.<br>h) Audit and surveillance: Monitor and audit user activity<br>- Track access and activity of roles and authorized users to detect and respond to anomalous behavior or incorrect use of privileges. |
| Ours-Secu-072 | Privileges should be assigned only when absolutely necessary, and accounts granted privileges should be identified and managed to minimize and monitor privileges.<br>a) Identify administrator and special privileges: Identifying which privileges within the cloud system correspond to administrator or special privileges. (Ex. creating and deleting instances, changing network configurations, managing user accounts, etc.)<br>b) Create and assign roles: Create new roles in Keystone and grant them administrative or special privileges. These roles are used to control access to cloud resources and services.<br>c) Assign a role to a user or group: Assign the role to a user or group with the permissions you identified. This allows you to grant administrative or special permissions to specific users or groups only. |
| | d) Set up access control and policies: Manage roles and permissions using Keystone's access control and policy features.<br>- Edit policy files to control administrator rights and special privileges, and restrict access to specific resources and services.<br>e) Maintain administrator and special permission lists: Keep separate lists of users or groups who have been granted special or administrator privileges.<br>- Track and manage which users or groups have which permissions.<br>f) Regular review and update: The list of users or groups with administrator rights and special privileges should be reviewed and updated regularly.<br>- Remove unnecessary permissions or reflect changed permissions.<br>g) Auditing and surveillance: Monitor and audit activities that use administrator and special privileges.<br>- Detect and respond to anomalous behavior or incorrect use of permissions. |
| Ours-Secu-073 | How to temporarily grant access to outsiders in a cloud system and properly dispose of them after their use is over should enhance security and prevent unnecessary access. You can manage access rights in the following ways.<br>- Just-in-time (JIT) access control: Implement JIT access control to ensure that outsiders only have access for as long as they need it. Access is only granted while performing the required task, and access is automatically revoked when the task is complete.<br>- Disposable accounts: Create and provide single-use accounts to outsiders. The account is valid for a single use or limited time only, and is automatically deactivated or deleted when the use is complete.<br>- Auto-stop or delete: Set a policy to automatically suspend or delete an account after a period of inactivity after it has been used by an outsider. This improves security by eliminating unnecessary accounts.<br>- Request administrator approval: Access to each account is carefully managed by requiring administrator approval whenever an outsider needs access to an account, limiting unnecessary access.<br>- Automatic audits and alerts: Monitor outsider activity and take action, such as automatically sending alerts or restricting access, when illegal activity or anomalous behavior is detected. |
| Ours-Secu-074 | In order to manage access to cloud systems and critical information, the appropriateness of granting access, utilization (long-term inactivity), and changes (retirement and leave of absence, job change, department change) should be checked regularly. Periodic reviews should be conducted to ensure that assigned accounts and access privileges are appropriate.<br>a) Establish review criteria and policies: Define criteria and policies for reviewing access privileges to clarify which privileges should be reviewed and under what conditions, which users or roles should be reviewed, etc.<br>b) Assign reviewers and roles: Specify who is responsible for reviewing access permissions. Typically, security admins, operations teams, compliance officers, etc. can perform reviews.<br>c)Determine review frequency: Determine how often you want to conduct regular reviews. The frequency can be monthly, quarterly, etc. and is based on the criticality of the system and your resources. |
| | d) Select for review: Select what to review. Prioritize high-value or sensitive permissions, etc.<br>e) Select review methods and tools: Choose how and what tools you'll use to conduct reviews. Utilize manual reviews, automated scripts, security information and event management systems, and more to conduct reviews.<br>f) Review process and execution: Run reviews on a set frequency<br>- Check permissions and roles, identify and fix unnecessary permissions, over-authorization, etc.<br>g) Document and report findings: Document review findings and report to relevant stakeholders; record what permissions were reviewed, what actions were taken, etc.<br>h) Actions and remediation: Perform any necessary actions and remediation based on the results of the review. Remove unnecessary permissions, adjust resource access, etc. |
| Ours-Secu-075 | Access control should be performed using attribute information (user's attributes, device information, location information) through an attribute-based access control model called ABAC.<br>a) To apply ABAC, first define the attributes that will control access to resources: user role, department, project, location, etc. Kestone's attributes include entity name, domain_id, parent_id, password, project id, enabled_service, etc.<br>b) Define access rules and policies for each resource and service (e.g., define a policy that only users with a "Department" attribute of "Dev Team" and a "Role" of "Admin" can access certain resources).<br>c) Use Keystone's policy engine to process actual access requests by matching attributes to policies. Keystone's policy engine determines whether a user or group is allowed access based on attributes.<br>d) Modify the "Policy Configuration File" within Keystone to enable ABAC policies and set the properties and policies to use. Allow you to add or change ABAC policies. You should test and validate that the ABAC policies you define are working properly. Use the Keystone API to try access control with your properties and policies applied, and verify that it works as expected.<br>e) Implement auditing and monitoring of the ABAC system's access rule evaluation and execution to detect and respond to anomalies and ensure that the ABAC rule to automatically apply rules when necessary. |
| Ours-Secu-076 | Entity and Activity Auditing should be applied to monitor for unusual or suspicious behavior to prevent evasive advanced threats, such as bypassing access restrictions.<br>a) Control access to Nova APIs and tightly configure policies for authorized users and roles.<br>- Tighten access restrictions by granting only necessary permissions and removing unnecessary ones.<br>b) Attribute-based access control (ABAC) must be implemented to enforce access control.<br>- User and role information: You need information about users and the roles they have.<br>- Identifiers and associated attributes must be defined for each user and role.<br>- Uniquely identify resources in Nova with UUIDs.<br>- Resource and service information: Information about the resources and services associated with a virtual machine (instance) in Nova. Defines an identifier and associated properties for each resource.<br>- Define policies and rules: Define which users can access certain resources or services when they meet certain conditions. |
| | - Manage properties and permissions: You need a way to define and manage the properties of users and resources. These properties are used in policies to determine a user's access rights.<br>c) Access rules and evaluation engine: Need an engine to define access rules and evaluate them. Determine and enforce access based on user attributes and policies.<br>d) Orchestration and automation: Implement orchestration and automation capabilities for managing and updating ABAC rules and automated enforcement of policies.<br>e) Data validation and input filtering: Validate data entered and perform filtering on inputs to prevent malicious input.<br>f) Audit and log monitoring: Apply Entity and Activity Auditing to audit access and activity to the Nova API and monitor logs to detect and respond to anomalous activity.<br>- Analyze logs to detect attempts to bypass access restrictions.<br>g) System and patch management: Keep systems related to Nova services up to date and regularly apply security patches and updates to address vulnerabilities through bypassing access restrictions.<br>h) Security review and testing: Regularly review code and settings related to the Nova API and perform security testing to identify and remediate vulnerabilities to bypassing access restrictions.<br>i) System surveillance and anomaly detection: Deploy a system to monitor Nova services and detect anomalies through SIEM to detect and respond to anomalous activity, such as bypassing access restrictions. |
| Ours-Secu-077 | OpenStack Glance should introduce or support technologies to protect. virtualized environments from malware such as viruses, worms, and Trojans.<br>a) Inspect and scan images: perform a malware scan before uploading images to Glance.<br>- Scan the files in the image for malware and block the upload if necessary.<br>b) Image encryption: Encrypt images as they are created to prevent leaks.<br>c) Image validation: Validate images and ensure they come from a trusted source.<br>d) Network security: Block malicious activity by implementing a SIEM to monitor network traffic and detect anomalies within your virtual environment.<br>e) Set up security groups and firewalls: Set up security groups and firewall policies for your virtual environment to control communication with the outside world and block malicious activity.<br>f) Update and patch: Regularly update and patch software related to the images and virtual environments you use to address known vulnerabilities.<br>g) Monitoring and logging: Monitor and log image usage and activity within your virtual environment to quickly detect and respond to malicious activity.<br>h) Backup and recovery: Establish a regular backup and recovery plan to allow for quick recovery in the event of a malware impact. |
| Ours-Secu-078 | To prevent API exploitation in OpenStack Neutron, the following safeguards should be put in place.<br>a) Access control and permission management: Minimize access to the Neutron API, granting only the minimum necessary permissions per user.<br>- Prevent unnecessary access by granting API access to only those who need it. |
| | b) Encryption: Encrypt data sent through the API to prevent credentials or sensitive information from being stolen.<br>- Use HTTPS to ensure the confidentiality of your data.<br>c) Harden API security: Adopt a security solution, such as a Web Application Firewall (WAF) or API Gateway, to protect API communications to detect and block malicious behavior or access.<br>d) Monitoring and auditing with SIEM: Monitor API access and activity, and detect anomalous behavior or illegal access attempts.<br>e) Authentication hardening: Implement strong authentication and authorization mechanisms for API access to prevent unauthorized access, even if credentials are revealed.<br>f) API token management: Securely manage tokens for API access, and set expiration times for tokens to restrict access if necessary.<br>g) Security updates and patches: Regularly apply security updates and patches to software and components related to OpenStack Neutron to address known vulnerabilities. |
| Ours-Secu-079 | Rather than using a static encryption key, you should use a secure protocol.<br>a) Neutron can communicate with Nova and Keystone via the following protocols<br>- TLS/SSL (Transport Layer Security/Secure Sockets Layer): Protocols primarily used to enhance the security of network communications, even in a Zero Trust environment. Encrypts communications between clients and Neutron servers to prevent man-in-the-middle attacks and data leakage. We recommend using TLS 1.3 or higher level protocols.<br>- Internet Protocol Security (IPsec): A protocol used to provide security at the network level, even in a Zero Trust environment.<br>Supports virtual private network (VPN) configuration and data encryption<br>- Secure Shell (SSH): A protocol used to secure remote logins and data transfers, even in a zero-trust environment. Encrypts communication between the server and user devices.<br>- WPA3 (Wi-Fi Protected Access 3): A protocol for enhancing the security of Wi-Fi networks; can be used to enhance the security of wireless networks within cloud environments.<br>- Hypertext Transfer Protocol Secure (HTTPS): Encrypts communication between websites to ensure the safety of data; used to enhance the security of web communications, even in zero-trust environments.<br>- SFTP (Secure File Transfer Protocol): A protocol for securing file transfers; can protect files containing information about Neutron Data Store's subnets, routers, ports, and IPs. |
| Ours-Secu-080 | You need to assign an Openstack Swift identifier so that only authorized users can access the internal network.<br>a) Project and user management: Create and manage projects (tenants) and users with OpenStack Keystone.<br>- Users have access to images and organize image-related resources through projects.<br>b) Define permissions and roles: Keystone uses roles to assign permissions to users. Define the permissions and roles needed for image-related tasks and assign them to users or projects.<br>c) Configuring access control policies: Configure access control policies to control access to Swift services.<br>- Associated with roles defined in Keystone, defining which users or projects can see which images. |
| | d) Control Glance API permissions: To control access to a Swift API server, set which users or projects can accept requests from that API server.<br>- Work with Keystone to authenticate and authorize users.<br>e) Management and auditing: Once you have implemented access control, you can utilize audit logs to monitor and review all access to your system. |

# 4 ANALYZE THE SECURITY OF EXISTING COMMERCIAL CLOUD SERVICE

In Chapter 4, we describe the security analysis results for cloud services with zero trust applied based on NIST, DoD, and the security requirements we derived to confirm the suitability and excellence of the security requirements we derived. We selected Microsoft Azure, Amazon Web Service, and Google Cloud as cloud services with Zero Trust because they have a high share in the cloud market[42]. By referring to the technical documentation and guidelines of Microsoft Azure[43], Amazon Web Service[44], and Google Cloud[45], and by using actual cloud services, we identified 158, 247, and 146 features for these three cloud services, respectively. In Section 4.1, we verify that Microsoft Azure, Amazon Web Services, and Google Cloud satisfy the security requirements identified by NIST, DoD, and our research team.

## 4.1 Results of the Security Analysis

As mentioned in Section 3.5 above, the security requirements derived by our research team are Zero Trust-based security requirements. To perform the security analysis on cloud services, we checked whether Microsoft Azure, Amazon Web Service, and Google Cloud are Zero Trust-based cloud services. We found that each cloud service meets all seven of NIST's Zero Trust Principles. This means that Microsoft Azure, Amazon Web Service, and Google Cloud are Zero Trust-based cloud services. For each service, we performed a security analysis based on the security requirements derived from NIST, DoD, and our research team. Before conducting the security analysis, we identified 25 and 52 security requirements from NIST SP 800-207 and DoD Zero Trust Reference Architecture, respectively. For NIST SP 800-207, we identified the requirements based on Chapter 3, "Logical Components of a Zero Trust Architecture," which contains strategies and network requirements for implementing Zero Trust[6]. For the DoD Zero Trust Reference Architecture, we identified based on Chapter 3, "Functional Taxonomy," which describes the technologies and capabilities required to implement Zero Trust[7]. The security analysis was conducted by analyzing the technical documents and guidelines of each cloud service and the actual service functions. The results of the security analysis were marked with "●" when all security requirements were met, "◑" when partially met, and "-" when not met. As a result of the security analysis, Microsoft Azure, Amazon Web Service, and Google Cloud were found to meet all 25 of NIST's requirements and 52 of DoD's requirements. However, in our security analysis, each cloud service failed to meet some of the security requirements we identified. This means that even if all the NIST and DoD security requirements are met, there are many residual threats. Furthermore, as discussed in Section 3.5, the existing NIST and DoD security requirements suffer from insufficiently descriptive guidelines that many organizations follow. Cloud service providers and users may not understand the security requirements due to the lack of clarity. This can increase leakage of sensitive information, hacking, data corruption, and other security risks. The following Table 6. summarizes the results of the security analysis for Microsoft Azure, Amazon Web Service, and Google Cloud.

**Table 6. Result of Security Anaylsis**

| Cloud Service | NIST | DoD | Ours |
|---|---|---|---|
| Microsoft Azure | ● | ● | ◑(5% unsatisfied) |
| Amazon Web Service | ● | ● | ◑(5% unsatisfied) |
| Google Cloud | ● | ● | ◑(6.25% unsatisfied) |

Based on our analysis of each cloud service's technical documentation, user guidelines, and actual cloud services, we found that Microsoft Azure met 76 of the 80 security requirements we identified and failed to meet 4. Amazon Web Service met 76 of the 80 security requirements we identified and failed to meet 4. Google Cloud met 75 of the 80 security requirements we identified and failed to meet 5.

### 4.1.1 Risk Analysis

In this paper, a risk analysis was conducted to determine the degree of impact of each threat. In a remote work environment, it can be costly and time-consuming to accept all risks, so it is important to manage risks cost-effectively. In this paper, Microsoft's DREAD model was used to prioritize the risks[46]. Using the OWASP Top 10[47] for prioritization, DREAD calculates a risk score based on five factors: damage potential, reproducibility of the attack, exploitability, affected users, and discoverability of the vulnerability. The risk score uses a simple system of low (1), medium (2), and high (3), with the higher the combined score for each element of DREAD, the higher the prioritization. The results of DREAD performed in this study are shown in Table 7.

**Table 7. DREAD**

| Threats | Security Requirements | D | R | E | A | D | Score |
|---|---|---|---|---|---|---|---|
| User Device Modification of account information | Ours-Secu-008 | 3 | 2 | 2 | 3 | 2 | 13 |
| Sensitive data collection | Ours-Secu-032 | 2 | 3 | 2 | 3 | 2 | 11 |
| Elevation of privilege | Ours-Secu-035 | 2 | 2 | 2 | 2 | 2 | 10 |
| Access Control | Ours-Secu-062 | 2 | 2 | 3 | 2 | 2 | 9 |
| Abusing permissions | Ours-Secu-074 | 3 | 2 | 2 | 2 | 3 | 12 |
| API calls that reveal authentication credentials | Ours-Secu-077 | 2 | 2 | 2 | 2 | 1 | 9 |

The requirements that all three cloud services have in common are forced password change upon initial access to the cloud environment (Ours-Secu-008), certificate revocation procedures (Ours-Secu-032), and periodic review of access privileges (Ours-Secu-074). Table 8. below shows the security requirements identified by us that Microsoft Azure, Amazon Web Service, and Google Cloud commonly fail to meet.

**Table 8. Ours Requirements with Microsoft Azure, Amazon Web Service, Google Cloud**

| ID | Ours Requirements | MS Azure | AWS | Google Cloud |
|---|---|---|---|---|
| Ours-Secu-008 | The requirements of password management procedures should be applied when designing the system. ... | ◑ | ◑ | ◑ |
| | e) Forced change of password at first access to information system, masking at the time of password processing (input, change). | | | |
| Ours-Secu-032 | A certificate disposal procedure should be established to prevent leakage of critical information stored in the Nova-cert/object store.<br>a) Request and confirmation of certificate retirement: Send a certificate retirement request to the SSL/TLS certificate issuing authority. At this time, the private key of the certificate and the reason for discarding the certificate must be provided together.<br>…<br>g) Record the reason for certificate disposal and information: Record the reason for certificate disposal and detailed information so that it can be verified later if necessary. | ◑ | ◑ | ◑ |
| Ours-Secu-074 | In order to manage access to cloud systems and important information, the appropriateness of access rights, use (long-term use), and changes (retirement and leave of absence, job change, department change) should be regularly checked. Periodic review should be conducted to ensure that the allocated account and access rights are appropriate.<br>a) Establishment of review criteria and policies: Define criteria and policies for reviewing access rights Clarify which rights should be reviewed under what conditions and which users or roles should be reviewed.<br>…<br>c) Determining review cycles: Periodic review cycles can be set monthly, quarterly, etc., determined by system importance and resources.<br>d) Selection of review targets: Selection of review targets. Priority selection of rights of high importance or sensitivity, etc.<br>e) Select a review method and tool: Select the method and tool to perform the review. Use manual reviews, automated scripts, security information, and event management systems to conduct reviews.<br>… | ◑ | ◑ | ◑ |

## 4.2 Potential Threats and Mitigation measures for Commercial Cloud Services

According to the security analysis, Microsoft Azure, Amazon Web Service, and Google Cloud are lacking in the following areas: enforcing password changes upon initial access to the cloud environment (Ours-Secu-008), certificate revocation procedures (Ours-Secu-032), and periodic access privilege reviews (Ours-Secu-074). The security requirement related to Forced password change on first access to cloud environment (Ours-Secu-008) states that the password must be changed on first access to the cloud environment. Each cloud service has a password policy, including authentication at the first login, password expiration period, etc. However, this security requirement is not satisfied when a cloud service user accesses the cloud environment for the first time because the user can log in with the same password set during registration. If the user's password is not forcibly changed when the cloud service user first accesses the cloud environment, a malicious attacker can access the cloud environment without authorization through the user's account and perform activities such as installing malicious software and modifying data. If an attacker gains access to a cloud service through account takeover, it can cause security incidents such as stealing important data or interrupting services. According to an IBM survey, 20% of data breaches in Korea were initially penetrated using user credentials[48]. The security requirement related to certificate revocation procedures (Ours-Secu-032) states that when a cloud service user requests that a certificate be revoked, the cloud service user must provide a reason for revocation, which must be recorded and documented. Each cloud service provides the ability to create and revoke certificates, but does not require the cloud service user to provide a reason for revoking the certificate. In addition, each cloud service allows users to permanently delete a certificate by clicking the Delete Certificate button. Therefore, it is unlikely that each cloud service provides the ability to record and document why a certificate is being revoked. Phishing, a popular attack method, uses a fake website to steal your account information and personal data. Phishing sites often have SSL/TLS certificates, so it's hard to tell if they're phishing or not based on the presence or absence of a certificate, so certificate management is important because phishing sites are created quickly. With the rise of HTTPS websites, SSL/TLS certificates secure online communications, so a compromise of an SSL certificate's private key can jeopardize the secure communication guarantees of end-to-end encryption. Kowsik Guruswamy, CTO of security firm Menlo Security, says that in a recent survey of HTTPS websites, 47.1% had vulnerable server software, including outdated Apache, Drupal, and WordPress. Sixty-seven percent of non-browser traffic originates from SSL. Guruswamy says that "hosting phishing links or drive by downloads on SSL makes it easier for attacks to succeed because no one checks them"[49]. If you don't record the reason for certificate retirement, it's hard to track down security issues or breaches when they occur. The security requirement related to Periodic Access Privilege Review (Ours-Secu-074) states that periodic reviews should be performed to ensure that cloud service users' accounts and access privileges are appropriate. Each cloud service performs access authorization control through RBAC or ABAC. However, there is no function to periodically review access rights, which does not meet the security requirements. Without periodic access review, users may continue to have privileges that are currently unnecessary. According to Palo Alto Networks, an analysis found that 99% of cloud users, roles, services, and resources were granted "excess privileges" that had not been used in 60 days[50]. The report warns that hackers can exploit these privileges to expand their attack radius both laterally and longitudinally. Insufficient access management also allows malicious actors disguising as normal users to read, modify, delete, etc. data in transit, which means they can gain unauthorized access to data and leak sensitive information about a company or organization. Some of the security requirements that Azure, AWS, and Google Cloud cannot meet are functional, while others are organizational, such as those related to access review. Therefore, in order to build a secure remote work environment based on Zero Trust, it is necessary to consider not only the functional aspects of Azure, AWS, and Google Cloud, but also the security education of users. Table 9. below shows the common unsatisfied items in Azure, AWS, and Google Cloud among the security requirements identified by us, along with possible security threats and mitigation measures.

**Table 9. Common possible security threats and mitigation measures**

| No. | Ours Requirements | Security Threats | Mitigation measures |
|---|---|---|---|
| Ours-Secu-008 | Forced change of password on first access to information systems. | - Malicious activity<br>- Misuse of authority<br>- Failure to comply with the latest security policies | Provide random passwords to users when accessing information systems for the first time |
| Ours- | Record the reason for the certificate | - Forgery of certification | Document the reasons for discarding the |
| Secu-032 | disposal and detailed information so that it can be verified later if necessary. | - Data integrity issues<br>- Difficulty in detecting security incidents | certificate and its contents when discard certificate. |
| Ours-Secu-074 | Periodic review should be conducted to ensure that the allocated account and access control are appropriate. | - Authorization Error<br>- Data leakage<br>- Increased likelihood of abuse | Periodically conducting reviews of permissions at the organization level. |

### 4.2.1 Potential threats and Mitigation measures in Microsoft Azure

In addition to the security requirements described in Sections 4.1 and 4.2 above, Microsoft Azure lacks security requirements related to installing software with clear provenance when establishing a virtual environment (Ours-Secu-035). Microsoft Azure provides regular security updates and monitoring features to detect and block malicious software, but it does not provide policies and guidelines that recommend cloud users to install only provenance-clear software. Without policies and guidelines to encourage the use of provenance, cloud service users are more likely to install unsafe software. This attack, which disguises itself as a legitimate file to trigger installation, can infect many software using organizations. Recently, the North Korean hacking group Andariel has been using certain asset management programs to distribute malware[51]. The group is known to use spear-phishing attacks, watering hole attacks, and software vulnerabilities for initial penetration, and it is also known to exploit other vulnerabilities to distribute malware during the attack process. As a result, users should be wary of files and software from unknown sources. This can compromise the confidentiality, integrity, and availability of assets on the system. As a result, organizations should conduct training on proper software installation in virtual environments and establish relevant policies and guidelines within the enterprise. Table 10. below shows potential threats in Microsoft Azure and how to mitigate them.

**Table 10. Possible security threats and mitigation measures of Microsoft Azure**

| No. | Ours Requirements | Security Threats | Mitigation measures |
|---|---|---|---|
| Ours-Secu-035 | Establish policies and guidelines that encourage cloud users to install only software with a clear source | - Installing Malicious Software<br>- Security attributes such as confidentiality, integrity, and availability are threatened | Organizations should establish policies and guidelines that encourage cloud users to install only software with a clear source. |

### 4.2.2 Potential threats and Mitigation measures in Amazon Web Services

In addition to the security requirements described in Sections 4.1 through 4.2 above, Amazon Web Services lacks security requirements related to user training on virtualization (Ours-Secu-062). AWS provides user training and user guidelines for various features of its cloud, but there are no features related to user training (Ours-Secu-062) for the storage space of an instance. Failure to comply with this security requirement could result in inefficient resource usage if users do not properly manage their instances or storage space. If a user overuses an instance or storage space, resources are depleted and normal service usage becomes difficult. An attacker can cause a targeted cloud service to overconsume limited system resources such as processor, memory, disk space, or network bandwidth, slowing down the system and making the service inaccessible to legitimate users. According to the Verizon Data Breach Investigations Report (DBIR), DDoS attacks are one of the biggest threats, accounting for 46% of all cybersecurity threats, with a high and growing trend[52].

Therefore, proper education and training should be provided to cloud service users within the organization. In addition, cloud service users should clearly establish policies and security rules for the storage of instances and operations related to virtual machines. Table 11. below shows potential threats in Amazon Web Service and how to mitigate them.

**Table 11. Possible security threats and mitigation measures of Amazon Web Service**

| No. | Ours Requirements | Security Threats | Mitigation measures |
|---|---|---|---|
| Ours-Secu-062 | Educate users and enhance their awareness of virtual environments | -Performance degradation due to inefficient resource use<br>- Denial of Service Attack (DoS) | Recognize and educate users about policies and security rules for instance and storage-related tasks |

### 4.2.3 Potential threats and Mitigation measures in Google Cloud

In addition to the security requirements described in Sections 4.1 through 4.2 above, Google Cloud lacks security requirements related to user training on virtualization (Ours-Secu-062) and image validation (Ours-Secu-077). As described in Section 4.2.2 above, the requirement for user training on virtual environment settings is an organizational requirement. The threats posed by non-compliance with this security requirement are the same as in Section 4.2.2 above. Google Cloud does not validate images when they are created by cloud service users, although if an image is in an unsupported format, it will not be created. An attacker could inject malicious code into the system through the image and infect the system. In addition, unvalidated images can be tampered with in the middle, and attackers can use them to exploit vulnerabilities in the system. The hacker group TeamTNT created a new Docker Hub account and uploaded malicious Docker images. They deployed malware by uploading malicious Docker images to Docker Hub that were disguised as legitimate[53]. To prevent and minimize these security threats, images should be validated when used in a cloud environment. Additionally, organizations within the enterprise should establish policies to ensure that only trusted images are used and that cloud users adhere to those policies. Table 12. below shows the potential threats in Google Cloud and how they can be mitigated.

**Table 12. Possible security threats and mitigations measures of Google Cloud**

| No. | Ours Requirements | Security Threats | Mitigation measures |
|---|---|---|---|
| Ours-Secu-062 | Educate users and enhance their awareness of virtual environments | -Performance degradation due to inefficient resource use<br>- Denial of Service Attack (DoS) | Recognize and educate users about policies and security rules for instance and storage-related tasks |
| Ours-Secu-077 | Validate an image and verify that it is from a trusted source | - Inserting malicious code<br>- Tampering image<br>- Vulnerability Exploits | Validates images when using them in a cloud environment, establishes and complies with policies to use only trusted images |

## 5. CONCLUSTION

We conducted a study to analyze security requirements for building a zero-trust-based remote work environment. Existing NIST and DoD security requirements are written in the abstract, making it difficult for service providers to identify and mitigate specific vulnerabilities. This can make it difficult for service providers to respond to security threats. In addition, the presence of unidentified vulnerabilities in a remote work environment can lead to security issues such as data leakage, unauthorized access, etc. Therefore, we performed threat modeling on OpenStack to derive detailed security requirements. As a result of the threat modeling, a total of 80 security requirements were derived, and security analysis was conducted on three commercial cloud services with zero trust principles. As a result of the security analysis, we found that Microsoft Azure, Amazon Web Service, and Google Cloud satisfy all existing NIST and DoD security requirements. However, Microsoft Azure, Amazon Web Service, and Google Cloud did not meet some of the specific requirements we identified. These results indicate that issues such as attacker malicious behavior, privilege misuse, and data leakage may occur within a remote work environment. Therefore, cloud service providers can implement and manage various security technologies such as data encryption, network security, and virtual environment security with more detailed guidelines through the security requirements presented in this paper. The security analysis shows that the role of cloud users as well as cloud service providers is important. Therefore, future research should be conducted to identify user considerations so that cloud users can utilize security-related policies and technologies. This is because organizations can establish their own security policies and utilize various security features provided by cloud service providers to maintain the safety of their own systems and securely manage cloud services. In light of these considerations, we believe that the detailed security requirements proposed in this paper can be used to design and develop a more secure zero-trust-based remote work environment.